\documentclass[12pt]{article}
\usepackage[english]{babel}
\usepackage{upquote}
\usepackage{subfig}

\usepackage{amsfonts,amssymb,amsmath, epsfig}
\usepackage{color,graphicx,graphics,psfrag}
\usepackage{amsmath,amstext,amssymb,amsfonts, amscd}
\usepackage[title]{appendix}
\usepackage{hyperref}

\def\Box{\leavevmode\vbox{\hrule
     \hbox{\vrule\kern4pt\vbox{\kern4pt}%
           \vrule}\hrule}}
\def\blackbox{\leavevmode\vrule height 5pt width 4pt depth 0pt\relax}
\def\endproof{\null\hfill {$\blackbox$}\bigskip}

\def\paragraph#1{{\bf #1\ }}

\newtheorem{lemma}{Lemma}[section]

\newtheorem{theorem}[lemma]{Theorem}

\newtheorem{definition}[lemma]{Definition}

\newtheorem{proposition}[lemma]{Proposition}

\newtheorem{remark}{Remark}[section]

\def\<{\langle}
\def\>{\rangle}

\def \R {{\mathbb R}}

\def \S {{\mathbb S}}

\def \ep {\varepsilon}

\def \o {\omega}
\def \Om{\Omega}

\title{Macroscopic models of collective motion with repulsion}
\author{Pierre Degond$^{1}$, Giacomo Dimarco$^{2}$, Thi Bich Ngoc Mac$^{3,4}$, Nan Wang $^{5} $ }
\date{}
\begin{document}

\maketitle

\begin{center}
1- Department of Mathematics, Imperial college London, \\
London SW7 2AZ, United Kingdom.\\
email:pdegond@imperial.ac.uk\\
$\mbox{}$ \\
2-Department of Mathematics, University of Ferrara, 44100 Ferrara, Italy\\
email: giacomo.dimarco@unife.it \\
$\mbox{}$ \\
3-Universit\'e de Toulouse; UPS, INSA, UT1, UTM ;\\
Institut de Math\'ematiques de Toulouse ; \\
F-31062 Toulouse, France. \\
$\mbox{}$ \\
4-CNRS; Institut de Math\'ematiques de Toulouse UMR 5219 ;\\
F-31062 Toulouse, France.\\
email: thi-bich-ngoc.mac@math.univ-toulouse.fr  \\
$\mbox{}$ \\
5-National University of Singapore, Department of Mathematics,\\
Lower Kent Ridge Road, Singapore 119076.\\
email: g0901252@nus.edu.sg \\
\end{center}

\vspace{0.5 cm}
\begin{abstract}

We study a system of self-propelled particles which interact with their neighbors via alignment and repulsion. The particle velocities result from self-propulsion and repulsion by close neighbors. The direction of self-propulsion is continuously aligned to that of the neighbors, up to some noise.  A continuum model is derived starting from a mean-field kinetic description of the particle system. It leads to a set of non conservative hydrodynamic equations. We provide a numerical validation of the continuum model by comparison with the particle model. We also provide comparisons with other self-propelled particle models with alignment and repulsion. 
\end{abstract}

\medskip
\noindent
{\bf Acknowledgements:} PD is on leave from CNRS, Institut de Math\'ematiques de Toulouse, France, and GD is on leave from Universit\'e Paul Sabatier, Institut de Math\'ematiques de Toulouse, France, where part of this research was conducted. TBNM wishes to thank the hospitality of the Department of Mathematics, Imperial College London for its hospitality. NW wishes to thank the Universit\'e Paul Sabatier, Institut de Math\'ematiques de Toulouse, France, for its hospitality. This work has been supported by the French 'Agence Nationale pour la Recherche (ANR)' in the frame of the contract 'MOTIMO' (ANR-11-MONU-009-01).


\vskip 0.4cm

\setcounter{equation}{0}
\section{Introduction}
\label{sec:intro}

The study of collective motion in systems consisting of a large number of agents, such as bird flocks, fish schools, suspensions of active swimmers (bacteria, sperm cells ), etc has triggered an intense literature in the recent years. We refer to \cite{Vicsek_Zafeiris_PhysRep12, Koch_FluidMech2011} for recent reviews on the subject. Many of such studies rely on a particle model or Individual Based Model (IBM) that describes the motion of each individual separately (see e.g in \cite{Aoki_BullJapSocSciFish92, Chate_etal_PRE08, Chuang_etal_PhysicaD07, Couzin_etal_JTB02, Cucker_Smale_IEEETransAutCont07, Henkes_etal_PRE11, Mogilner_etal_JMB03, Motsch_Tadmor_JSP11, Szabo_PRL06}).

In this work, we aim to describe dense suspensions of elongated self-propelled particles in a fluid, such as sperm. In such dense suspensions, steric repulsion is an essential ingredient of the dynamics. A large part of the literature is concerned with dilute suspensions \cite{Hernandez-Ortiz_etal_JPhysCondMat09, Koch_FluidMech2011, Pedley_etal_JFM88, Saintillan_Fluids08, Woodhouse_Goldstein_PRL12}. In these approaches, the Stokes equation for the fluid is coupled to the orientational distribution function of the self-propelled particles. However, these approaches are of ``mean-field type'' i.e. assume that particle interactions are mediated by the fluid through some kinds of averages. These approaches do not deal easily with short-range interactions such as steric repulsion or interactions mediated by lubrication forces. Additionally, these models assume a rather simple geometry of the swimmers, which are reduced to a force dipole, while the true geometry and motion of an actual swimmer, like a sperm cell, is considerably more complex. 

In a recent work \cite{Peruani_etal_PRE06}, Peruani et al showed that, for dense systems of elongated self-propelled particles, steric interaction results in alignment. Relying on this work, and owing to the fact that the description of swimmer interactions from first physical principles is by far too complex, we choose to replace the fluid-mediated interaction by a simple alignment interaction of Vicsek type \cite{Vicsek_etal_PRL95}. In the Vicsek model, the agents move with constant speed and attempt to align with their neighbors up to some noise. Many aspects of the Vicsek model have been studied, such as phase transitions \cite{Aldana_etal_PRL07, Chate_etal_PRE08, Degond_etal_JNonlinearSci13, Degond_etal_preprint13, Gretoire_Chate_PRL04, Vicsek_etal_PRL95}, numerical simulations \cite{Motsch_Navoret_MMS11}, derivation of macroscopic models \cite{Bertin_etal_JPA09, Degond_Motsch_M3AS08}.

The alignment interaction acting alone may trigger the formation of high particle concentrations.  However, in dense suspensions, volume exclusion prevent such high densities to occur. When distances between particles become too small, repulsive forces are generated by the fluid or by the direct reaction of the bodies one to each other. These forces contribute to repel the particles and to prevent further contacts. To model this behavior, we must add a repulsive force to the Vicsek alignment model. Inspired by \cite{Baskaran_Marchetti_PRL10, Henkes_etal_PRE11, Szabo_PRL06} we consider the possibility that the particle orientations (i.e the directions of the self-propulsion force) and the particle velocities may be different. Indeed, steric interaction may push the particles in a direction different from that of their self-propulsion force. 

We consider an overdamped regime in which the velocity is proportional to the force through a mobility coefficient. The overdamped limit is justified  by the fact that the background fluid is viscous and thus the forces due to friction are very large compared to those due to motion. Indeed, for micro size particles, the Reynolds number is very small ($\sim 10^{-4}$) and thus the effect of inertia can be neglected. Finally, differently from \cite{Baskaran_Marchetti_PRL10, Henkes_etal_PRE11, Szabo_PRL06} we consider an additional term describing the relaxation of the particle orientation towards the direction of the particle velocity. We also take into account a Brownian noise in the orientation dynamics of the particles. This noise may take into account the fluid turbulence for instance. Therefore, the particle dynamics 
results from an interplay between relaxation towards the mean orientation of the surrounding particles, relaxation towards the direction of the velocity vector and Brownian noise. From now on we refer to the above described model as the Vicsek model with repulsion.

Starting from the above described microscopic dynamical system we successively derive mean-field equations and hydrodynamic equations. Mean field equations are valid when the number of particles is large and describe the evolution of the one-particle distribution, i.e. the probability for a particle to have a given orientation and position at a given instant of time. Expressing that the spatio-temporal scales of interest are large compared to the agents' scales leads to a singular perturbation problem in the kinetic equation. Taking the hydrodynamic limit, (i.e. the limit of the singular perturbation parameter to zero) leads to the hydrodynamic model.  Hydrodynamic models are particularly well-suited to systems consisting of a large number of agents and to the observation of the system's large scale structures. Indeed, the computational cost of IBM increases dramatically with the number of agents, while that of hydrodynamic models is independent of it. With IBM, it is also sometimes quite cumbersome to access observables such as order parameters, while these quantities are usually directly encoded into the hydrodynamic equations.  

The derivation of hydrodynamic models has been intensely studied by many authors. Many of these models are based on phenomenological considerations \cite{Toner_etal_AnnPhys05} or derived from moment approaches and ad-hoc closure relations \cite{Baskaran_Marchetti_PRL10, Bertin_etal_JPA09, Ratushnaya_etal_PhysicaA07}. The first mathematical derivation of a hydrodynamic system for the Vicsek model has been proposed in \cite{Degond_Motsch_M3AS08}. We refer to this model to as the Self-Organized Hydrodynamic (SOH) model. One of the  main contributions of \cite{Degond_Motsch_M3AS08} is the concept of ``Generalized Collision Invariants'' (GCI) which permits the derivation of macroscopic equations for a particle system in spite of its lack of momentum conservation. The SOH model has been further refined in \cite{Degond_etal_MAA13, Frouvelle_M3AS12}. 

Performing the hydrodynamic limit in the kinetic equations associated to the Vicsek model with repulsion leads to the so-called ``Self-Organized Hydrodynamics with Repulsion'' (SOHR) system. The SOHR model consists of a continuity equation for the density $\rho$ and an evolution equation for the average orientation $\Om \in \S^{n-1}$ where $n$ indicates the spatial dimension. More precisely, the model reads
\begin{eqnarray}
& \partial_t\rho +  \nabla_x \cdot(\rho U )  = 0, \label{macro1_intro} \\
&  \rho \partial_t \Om +  \rho (V \cdot \nabla_x)\Om +  P_{\Om^\perp}\nabla_x p(\rho) =  \gamma P_{\Om^\perp} \Delta_x(\rho \Om),  \label{macro2_intro}	\\
& \vert \Om \vert = 1,	\label{macro3_intro}
\end{eqnarray}
where
\begin{eqnarray}
& U =  c_1v_0 \Om - \mu \Phi_0 \nabla_x \rho , \quad V =  c_2 v_0 \Om - \mu \Phi_0 \nabla_x \rho , \label{macro4_intro}\\
& p(\rho ) = v_0 d \rho  + \alpha \mu \Phi_0 \big( (n-1)d + c_2 \big) \dfrac{\rho^2}{2}, \quad \gamma = k_0 \big( (n-1)d + c_2\big). \label{macro5_intro}
\end{eqnarray}
The coefficients $c_1$, $c_2$, $v_0$, $\mu$, $\Phi_0$, $d$, $\alpha$, $k_0$ are associated to the microscopic dynamics and will be defined later on. The symbol $P_{\Omega^\bot}$ stands for the projection matrix $ P_{\Omega^\bot} = \mbox{Id} - \Omega \otimes \Omega$ of ${\mathbb R}^n$ on the hyperplane $\Omega^\bot$. The SOHR model is similar to the SOH model obtained in \cite{Degond_etal_MAA13}, but with several additional terms which are consequences of the repulsive force at the particle level. The repulsive force intensity is characterized by the parameter $\mu \Phi_0$. In the case $\mu \Phi_0 = 0$, the SOHR system is reduced to the SOH one. 

We first briefly describe the original SOH model. Inserting (\ref{macro4_intro}), (\ref{macro5_intro}) with $\mu \Phi_0 = 0$ into (\ref{macro1_intro}), (\ref{macro2_intro}) leads to  
\begin{eqnarray}
& \partial_t\rho +  c_1 v_0 \, \nabla_x \cdot(\rho \Omega )  = 0, \label{macro6_intro} \\
&  \rho \partial_t \Om +  c_2 v_0 \rho (\Omega \cdot \nabla_x)\Omega +  v_0 d \,  P_{\Om^\perp}\nabla_x \rho =  \gamma P_{\Om^\perp} \Delta_x(\rho \Om),  \label{macro7_intro}
\end{eqnarray}
together with (\ref{macro3_intro}). This model shares similarities with the isothermal compressible Navier-Stokes (NS) equations. Both models consist of a non linear hyperbolic part supplemented by a diffusion term. Eq. (\ref{macro6_intro}) expresses conservation of mass, while Eq. (\ref{macro7_intro}) is an equation for the mean orientation of the particles. It is not conservative, contrary to the corresponding momentum conservation equation in NS. The two equations are supplemented by the geometric constraint (\ref{macro3_intro}). This constraint is satisfied at all times, as soon as it is satisfied initially. The preservation of this property is guaranteed by the projection operator $P_{\Om^\perp}$. A second important difference between the SOH model and NS equations is that the convection velocities for the density and the orientation, $v_0 c_1$ and $v_0 c_2$ respectively are different while for NS they are equal. That $c_1 \not  = c_2$ is a consequence of the lack of Galilean invariance of the model (there is a preferred frame, which is that of the fluid). The main consequence is that the propagation of sound waves is anisotropic for this type of fluids \cite{Toner_etal_AnnPhys05}. 

The first main difference between the SOH and the SOHR system is the presence of the terms $\mu \Phi_0 \nabla_x \rho$ in the expressions of the velocities $U$ and $V$. Inserting this term in the density Eq. (\ref{macro1_intro}) results in a diffusion-like term $- \mu \Phi_0 \, \nabla_x \cdot(\rho ( \nabla_x \rho))$ which avoids the formation of high particle concentrations. This term shows similarities with the non-linear diffusion term in porous media models. Similarly, inserting the term $\mu \Phi_0 \nabla_x \rho$ in the orientation Eq.  (\ref{macro2_intro}) results in a convection term in the direction of the gradient of the density. Its effect is to force particles to change direction and move towards regions of lower concentration. The second main difference is the replacement of the linear (with respect to $\rho$) pressure term $v_0 d \,  P_{\Om^\perp}\nabla_x \rho$ by a nonlinear pressure $p(\rho)$ in the orientation Eq.  (\ref{macro2_intro}). The nonlinear part of the pressure enhances the effects of the repulsion forces when concentrations become high. 

To further establish the validity of the SOHR model (\ref{macro1_intro})-(\ref{macro5_intro}), we perform numerical simulations and compare them to those of the underlying IBM. To numerically solve the SOHR model, we adapt the relaxation method of \cite{Motsch_Navoret_MMS11}. In this method, the unit norm constraint (\ref{macro3_intro}) is  abandonned and replaced by a fully conservative hyperbolic model in which $\Omega$ is supposed to be in ${\mathbb R}^n$. However, at the end of each time step of this conservative model, the vector $\Omega$ is normalized. Motsch and Navoret showed that the relaxation method provides numerical solutions of the SOH model which are consistent with those of the particle model. The resolution of the conservative model can take advantage of the huge literature on the numerical resolution of hyperbolic conservation laws (here specifically, we use \cite{Degond_Peyrard_99}). We adapt the technique of \cite{Motsch_Navoret_MMS11} to include the diffusion fluxes. Using  these approximations, we numerically demonstrate the good convergence of the scheme for smooth initial data and the consistency of the solutions with those of the particle Vicsek model with repulsion. 

The outline of the paper is as follows. In section 2, we introduce the particle model, its mean field limit, the scaling and the hydrodynamic limit. In Section 3 we present the numerical discretization of the SOHR model, while in Section 4 we present several numerical tests for the macroscopic model and a comparison between the microscopic and microscopic models. Section 5 is devoted to draw a conclusion. Some technical proofs will be given in the Appendices.

\setcounter{equation}{0}
\section{ Model hierarchy and main results }
\label{sec:IBM}

\subsection{The individual based model and the mean field limit}

We consider a system of  $N$-particles each of which is described by its position $ X_k(t) \in \R^n$, its velocity $v_k(t) \in \R^n,$ and its direction $ \o_k(t) \in \S^{n-1} ,$ where $  k \in \{ 1, \cdots , N \}$, $n$ is the spatial dimension and $\S^{n-1}$ denotes the unit sphere.   The particle ensemble satisfies the following stochastic differential equations 
\begin{eqnarray} 
\dfrac{dX_k}{dt} & = &  v_k,
\label{overdamped1}\\
v_k	& = & v_0 \, \o_k  -  \mu \, \nabla_x \Phi(X_k(t),t),
\label{overdamped2}\\
d\o_k	 & = & P_{\o_k^\perp}\circ (\nu \, \bar{\o}(X_k(t),t) \, dt  + \alpha \, v_k \, dt + \sqrt{2D} \, dB^k_t ).\label{overdamped3}
\end{eqnarray}	
Eq. (\ref{overdamped1}) simply expresses the spatial motion of a particle of velocity $v_k$. Eq (\ref{overdamped2}) shows that the velocity  $v_k$ is composed of two components:  a self-propulsion velocity of constant magnitude $v_0$ in direction $\o_k$ and a repulsive force proportional to the gradient of a potential $\Phi(x,t)$ with mobility coefficient $\mu$.  Equation (\ref{overdamped3}) describes the time evolution of the orientation. The first term models the relaxation of the particle orientation towards the average orientation $\bar{\o}(X_k(t),t)$ of its neighbors with rate $\nu$.  The second term models the relaxation of the particle orientation towards the direction of the particle velocity $v_k$ with rate $\alpha$.  Finally, the last term describes standard independent white noises $dB^k_t$ of intensity $\sqrt{2D}$. The symbol $\circ$ reminds that the equation has to be understood in the Stratonovich sense. Under this condition and thanks to the presence of $P_{\o^\perp}$, the orthogonal projection onto the plane orthogonal to $\o$  (i.e  $P_{\o^\perp} = (\mbox{Id} - \o \otimes \o),$ where $\otimes $ denotes the tensor product of two vectors and $\mbox{Id}$  is the identity matrix), the orientation $\omega_k$ remains on the unit sphere. We assume that $v_0$, $\mu$, $\nu$ $\alpha$, $D$ are strictly positive constants. 

The repulsive potential $\Phi(x,t)$ is the resultant of binary interactions mediated by the binary interaction potential $ \phi $. It is given by: 
\begin{equation}
\Phi(x,t) = \dfrac{1}{N} \sum_{i=1}^{N} \nabla \phi \Big( \frac{|x - X_i|}{r} \Big)
\label{eq:pot}
\end{equation}
where the binary repulsion potential $\phi(|x|)$ only depends on the distance. We suppose that $x \mapsto \phi(|x|)$ is smooth (in particular implying that  $\phi'(0) =0$ where the prime denotes the derivative with respect to $|x|$). We also suppose that 
$$ \phi \geq 0, \qquad \int_{{\mathbb R}^n} \phi(|x|) \, dx < \infty, $$
in particular implying that $\phi(|x|) \to 0$ as $ |x| \rightarrow  \infty$. The quantity $r$ denotes the typical range of $\phi$ and the fact that $\phi \geq 0$ ensures that $\phi$ is repulsive. In the numerical test Section, we will propose precise expressions for this potential force. 

The mean orientation $\bar{\o}(x,t)$ is defined by
\begin{equation}
\bar{\o}(x,t) =  \dfrac{{\mathcal J}(x,t)}{\vert {\mathcal J}(x,t) \vert},  \qquad   {\mathcal J} (x,t)  = \frac{1}{N} \sum\limits_{i=1}^{N} K\Big(\frac{|x-X_i|}{R}\Big) \o_i.
\label{eq:meanorient}
\end{equation}	
It is constructed as the normalization of the vector $ {\mathcal J} (x,t)$ which sums up all orientation vectors $\omega_i$ of all the particles which belong to the range of the ``influence kernel'' $K(|x|)$.  The quantity $R>0$ is the typical range of the influence kernel $K (|x|/R)$, which is supposed to depend only on the distance.  It measures how the mean orientation at the origin is influenced by particles at position $x$. Here, we assume that $x \to K(|x|)$ is smooth at the origin and compactly supported. For instance, if $K$ is the indicator function of the ball of radius $1$, the quantity  $\bar \omega (x,t)$ computes the mean direction of the particles which lie in the sphere of radius $R$ centered at $x$ at time $t$.

\begin{remark}
(i) In the absence of repulsive force (i.e. $\mu = 0$), the system reduces to the time continuous version of the Vicsek model proposed in \cite{Degond_Motsch_M3AS08}. \\
(ii) The model presented is the so called overdamped limit of the model consisting of (\ref{overdamped1}) and (\ref{overdamped3}) and where (\ref{overdamped2}) is replaced by: 
\begin{eqnarray} \label{nonoverdamped4}
\epsilon \dfrac{dv_k}{dt}	 & = & \lambda_1 (v_0 \o_k - v_k) - \lambda_2  \nabla_x \Phi(X_k(t),t).
\end{eqnarray}	
with $\mu = \lambda_2/\lambda_1.$ Taking the limit $ \epsilon \rightarrow 0 $ in (\ref{nonoverdamped4}), we obtain  (\ref{overdamped2}). As already mentioned in the Introduction, for microscopic swimmers, this limit is justified by the very small Reynolds number and the very small inertia of the particles.  
\end{remark}

We now introduce the mean field kinetic equation which describes the time evolution of the particle system in the large $N$ limit. The unknown here is the one particle distribution function $f(x, \o, t)$ which depends on the position $x \in \R^n$, orientation $\o\in \S^{n-1}$ and time $t$. The evolution of $f$ is governed by the following system
\begin{eqnarray}
& & \hspace{-1cm} \partial_t f +  \nabla_x  \cdot (v_f f) + \nu \, \nabla_\o \cdot( P_{\o^\perp}  \bar{\o}_f  f)  +  \alpha \, \nabla_\o \cdot (P_{\o^\perp} v_f  f ) -  D \Delta_\o f = 0,  
\label{eqmeanfield1}\\
& & \hspace{-1cm}  v_f(x, t)    =   	v_0 \o  - \mu \nabla_x \Phi_f(x,t),    
\label{eqmeanfield2}
\end{eqnarray}
where the repulsive potential and the average orientation are given by 	
\begin{eqnarray}
& & \Phi_f(x,t)    =   \int_{\S^{n-1} \times \R^n} \phi\left(\dfrac{|x-y|}{r}\right) \,f(y, w, t) \, dw \, dy,   
\label{mfeq1} \\
 & &\bar{\o}_f  (x,\o, t)      =    \dfrac{\mathcal J_f(x,t)}{|\mathcal J_f(x,t)|},   
\label{mfeq2}\\	     	
& & \mathcal J_f (x,t)    =  	\int_{\S^{n-1} \times \R^n} K\left(\dfrac{|x-y|}{R}\right) \, f(y, w, t) \, w \, dw \, dy 
\label{mfeq3}.		 
\end{eqnarray}	
Equation (\ref{eqmeanfield1}) is a Fokker-Planck type equation. The second term at the left-hand side of (\ref{eqmeanfield1}) describes particle transport in physical space with velocity $ v_f$ and is the kinetic counterpart of Eq. (\ref{overdamped1}). The third, fourth and fifth terms describe transport in orientation space and are the kinetic counterpart of Eq. (\ref{overdamped3}). The alignment interaction is expressed by the third term, while the relaxation force towards the velocity $v_f$ is expressed by the fourth term. The fifth term represents the diffusion due to the Brownian noise in orientation space. The projection $P_{\o^\perp}$ insures that the force terms are normal to $\o$. The symbols $\nabla_\o \cdot$ and $\Delta_\o$ respectively stand for the divergence of tangent vector fields to $\S^{n-1}$ and the Laplace-Beltrami operator on $\S^{n-1}$. Eq. (\ref{eqmeanfield2}) is the direct counterpart of (\ref{overdamped2}). 

Eq. (\ref{mfeq1}) is the continuous counterpart of Eq. (\ref{eq:pot}).  Indeed, letting $f$ be the empirical measure
$$f = \frac{1}{N} \sum_{i=1}^N \delta_{(x_i(t), \omega_i(t))} (x,\omega), $$
in (\ref{mfeq1}) (where $\delta_{(x_i(t), \omega_i(t))} (x,\omega)$ is the Dirac delta at $(x_i(t), \omega_i(t))$) leads to (\ref{eq:pot}). Similarly, Eqs. (\ref{mfeq2}), (\ref{mfeq3}) are the continuous counterparts of (\ref{eq:meanorient}) (by the same kind of argument). The rigorous convergence of the particle system to the above Fokker-Planck equation (\ref{eqmeanfield1}) is an open problem. We recall however that, the derivation of the kinetic equation for the Vicsek model without repulsion has been done in \cite{Bolley_etal_AML11} in a slightly modified context.

\subsection{Scaling}
	
In order to highlight the role of the various terms, we first write the system in dimensionless form. We chose $t_0$ as unit of time and choose 
$$ x_0 = v_0 t_0, \qquad f_0 = \frac{1}{x_0^n}, \qquad \phi_0 = \frac{v_0^2 \, t_0}{\mu}, $$
as units of space, distribution function and potential. We introduce the dimensionless variables:
$$
\tilde{x} = \dfrac{x}{x_0}, \qquad  \tilde{t} = \dfrac{t}{t_0}, \qquad \tilde{f} = \dfrac{f}{f_0}, \qquad \tilde{\phi} = \dfrac{\phi}{\phi_0}, 
$$
and the dimensionless parameters
$$
\breve R = \dfrac{R}{x_0}, \qquad  \breve r = \dfrac{r}{x_0},  \qquad  \breve D  = t_0 D, \qquad \breve \nu = t_0 \nu, \qquad  \breve \alpha = \alpha x_0.
$$
In the new set of variables $(\tilde x,\tilde t)$, Eq.  (\ref{eqmeanfield2}) becomes	(dropping the tildes and the $\breve{}$ for simplicity): 
$$
v_f =	 \o  - \nabla_x \Phi_f(x,t), 
$$
\noindent	
while $f$, $\Phi_f$, $\bar \omega_f$, ${\mathcal J}_f$are still given by (\ref{eqmeanfield2}), (\ref{mfeq1}), (\ref{mfeq2}), (\ref{mfeq3}) (now written in the new variables). 

We now define the regime we are interested in. We assume that the ranges $R$ and $r$ of the interaction kernels $K$ and $\phi $ are both small but with $R$ much larger than $r$. More specifically, we assume the existence of a small parameter $\varepsilon \ll 1$ such that:
$$ R = \sqrt{\ep} \hat{R}, \qquad r = \ep \hat{r} \qquad \mbox{ with } \qquad \hat{R}, \, \hat{r} = \mathcal O(1).$$ 
We also assume that the diffusion coefficient $D$ and the relaxation rate to the mean orientation $\nu$ are large and of the same orders of magnitude (i.e. $d = D/\nu = {\mathcal O}(1)$),  while the relaxation to the velocity $\alpha$ stays of order $1$, i.e.
$$ \nu = \dfrac{1}{\ep}, \qquad d = \dfrac{D}{\nu}  = {\mathcal O}(1), \qquad \alpha = {\mathcal O}(1). $$
With these new notations, dropping all 'hats', the distribution function $f^\ep(x,\o,t)$ (where the superscript $\varepsilon$ now higlights the dependence of $f$ upon the small parameter $\varepsilon$) satisfies the following Fokker-Plank equation 	
\begin{eqnarray}
&&\hspace{-1.2cm}   \ep \Big( \partial_t f^\ep +  \nabla_x  \cdot (v_{f^\varepsilon}^\ep f^\ep) \Big)+ \nabla_\o \cdot( P_{\o^\perp} \bar \o^\ep_{f^\varepsilon}  f^\ep) +  \ep \alpha \nabla_\o \cdot (P_{\o^\perp} v^\ep_{f^\varepsilon} f^\ep)  -   d \Delta_\o f^\ep = 0, 
\label{eqmeanfield5} \\
&&\hspace{-1cm}  v^\ep_f =	  \o  -  \nabla_x \Phi_f^\ep(x,t),	
\label{eqmeanfield6}
\end{eqnarray}
where the repulsive potential and the average orientation are now given by
\begin{eqnarray*}
& & \hspace{-1cm}    \Phi_f^\ep(x,t)    =    \int_{\S^{n-1} \times \R^n} \phi\Big(\dfrac{|x-y|}{\ep r } \Big) \, f^\ep(y, w, t) \, dw \, dy,    
\\
& & \hspace{-1cm} 	\bar{\o}^\ep_f   =  \dfrac{\mathcal J^\ep_{f}(x,t)}{|\mathcal J^\ep_f(x,t)|} ,    \quad \mathcal J^\ep_f (x,t)  =  	 \int_{\S^{n-1} \times \R^n} K\Big(\dfrac{|x-y|}{\sqrt{\ep}R } \Big)  \,  f^\ep (y, w, t) \, w \, dw \, dy 
\end{eqnarray*}

Now, by Taylor expansion and the fact that the kernels $K$, $\phi$ only depend on $|x|$, we obtain (provided that $K$ is normalized to $1$ i.e. $\int_{\mathbb R} K(|x|) \, dx = 1$) :
\begin{eqnarray}
& & \hspace{-1cm} v_f^\ep(x,t) =\o  - \Phi_0 \nabla_x \rho_f^\ep  +  \mathcal O(\ep^2),   \label{expansion1}\\
&  & \hspace{-1cm} \bar{\o}^\ep_f (x,t)  =  G_f^0(x,t) + \ep G^1_f(x,t) +  \mathcal O(\ep^2), \label{expansion2}\\
& & \hspace{-1cm}  G_f^0(x,t) =   \Om_f(x,t)  , \quad  G^1_f(x,t) =   \dfrac{k_0}{ | J_f|} P_{\Om_f^\perp} \Delta_x J_f, \nonumber
\end{eqnarray}
where the coefficients $k_0, \Phi_0$ are given by
\begin{equation}
\label{k0}
k_0 = \dfrac{R^2}{2n} \int_{x\in \R^n} K (|x|)|x|^2 dx , \quad \Phi_0 = \int_{x\in \R^n} \phi(x) dx .
\end{equation}
For example, if $K$ is the indicator function of the ball of radius $1,$ then $ k_0 = \vert \S^{n-1} \vert / 2n(n+2) ,$ where $\vert \S^{n-1} \vert $ is the volume of the sphere $\S^{n-1}$. In the cases $d = 2$ and $d = 3$, we respectively get $ k _0= \pi/8 $ and $ k_0 =2\pi/15 .$ The local density $\rho_f,$ the local current density $J_f$ and local average orientation $\Om_f$ are defined by
\begin{eqnarray}
& & \hspace{-1cm} \rho_f(x,t) =  \int_{\S^{n-1} } f(x, w, t) \, dw, \quad
\label{f1}\\	
& & \hspace{-1cm}  
J_f(x,t)  = \int_{\o \in \S^{n-1}} f(x, w,t)  \, w \, d w, \quad
\Om_f(x,t) = \dfrac{J_f(x,t)}{|J_f(x,t)|}. 
\label{f2} 
\end{eqnarray}
More details about this Taylor expansion are given in Appendix \ref{sec:taylorexpasion} .	Let us observe that this scaling, first proposed in \cite{Degond_etal_MAA13} is different from the one used in \cite{Degond_Motsch_M3AS08} and results in the appearance of the viscosity term at the right-hand side of Eq. (\ref{macro2_intro}). 

Finally, if we neglect the terms of order $\ep^2$ and we define the so-called collision operator $Q(f)$ by
$$
  Q(f)   =     - \nabla_\o \cdot (P_{\o^\perp}  \Om_f f) + d \Delta_\o f,
$$
the rescaled system (\ref{eqmeanfield5}), (\ref{eqmeanfield6}) can be rewritten as follows
\begin{eqnarray}
&&\hspace{-1cm} \ep \Big ( \partial_t f^\ep +  \nabla_x  \cdot (v^\ep_f f^\ep) + \alpha \, \nabla_\o \cdot (P_{\o^\perp} v^\ep_f f^\ep) + \nabla_\o \cdot (P_{\o^\perp}  G^1_{f^\ep} f^\ep) \Big ) =  Q(f^\ep),  
\label{lastscaling1}\\
&&\hspace{-1cm}  v_{f^\ep}(x, \o,t) = \o  - \Phi_0 \nabla_x \rho_{f^\ep} , \quad  G^1_{f^\ep}(x,t)  =  \dfrac{k_0}{ | J_{f^\ep}|} P_{\Om_f^\perp} \Delta_x J_{f^\ep}  
\label{lastscaling2}
	\end{eqnarray}

\subsection{Hydrodynamic limit}

The aim is now to derive a hydrodynamic model by taking the limit $\ep \rightarrow 0$ of system (\ref{lastscaling1}), (\ref{lastscaling2}) where the local density $\rho_f,$ the local current $J_f$ and the local average orientation $\Om_f$ are defined by (\ref{f1}), (\ref{f2}). 

We first introduce the von Mises-Fisher (VMF) probability distribution $M_{\Omega}(\omega)$ of orientation $\Omega \in {\mathbb S}^{n-1}$ defined for $\omega \in {\mathbb S}^{n-1}$ by: 
$$
M_\Om(\o)=Z^{-1} \,  \exp\left(\dfrac{\o \cdot \Om}{d}\right),  \qquad Z = \int_{\omega \in {\mathbb S}^{n-1}} \exp\left(\dfrac{\o \cdot \Om}{d}\right) \, d\omega
$$
An important parameter will be the flux of the VMF distribution, i.e. $\int_{\o \in \S^{n-1}} M_\Om(\o) \o d\o$. By obvious symmetry consideration, we have 
$$
\int_{\o \in \S^{n-1}} M_\Om(\o) \, \o \, d\o = c_1 \Om, 
$$
where the quantity $c_1 = c_1(d)$ does not depend on $\Om$, is such that $0 \leq c_1(d) \leq 1$ and is given by 
\begin{equation}
\label{c1}
c_1(d) = \int_{\o \in \S^{n-1}} M_\Om(\o) \, (\o \cdot \Omega) \, d\o.
\end{equation}
When $d$ is small, $M_\Omega$ is close to a Dirac delta $\delta_\Omega$ and represents a distribution of perfectly aligned particles in the direction of $\Omega$. When $d$ is large, $M_\Omega$ is close to a uniform distribution on the sphere and represents a distribution of almost totally disordered orientations. 
The function $d \in {\mathbb R}_+ \mapsto c_1(d) \in [0,1]$ is strictly decreasing with $\lim_{d \to 0} c_1(d) = 1$,  $\lim_{d \to \infty} c_1(d) = 0$. Therefore, $c_1(d)$ represents an order parameter, which corresponds to perfect disorder when it is close to $0$ and perfect alignment order when it is close to $1$. 

\medskip
\noindent
We have following theorem:

\begin{theorem}
\label{thm:sm}
Let $f^\ep$ be the solution of (\ref{lastscaling1}), (\ref{lastscaling2}). Assume that there exists $f$ such that 
\begin{equation}
f^\ep \rightarrow f \quad \mbox{ as } \quad \ep \rightarrow 0, 
\label{eq:converg}
\end{equation} 
pointwise as well as all its derivatives. Then, there exist $ \rho(x,t)$ and $\Om(x,t)$ such that
	\begin{equation}\label{f0}
		f(x,\o,t) = \rho(x,t) M_{\Om(x,t)}(\o),
	\end{equation}			
Moreover, the functions $\rho(x,t), \Om(x,t)$ satisfy the  following equations\begin{eqnarray}
&&\hspace{-1cm} \partial_t\rho +  \nabla_x \cdot(\rho U )  = 0, 
\label{macro1} \\
&&\hspace{-1cm}  \rho \big(\partial_t \Om +  (V \cdot \nabla_x)\Om \big) +  P_{\Om^\perp}\nabla_x p(\rho) =  \gamma P_{\Om^\perp} \Delta_x(\rho \Om),   
\label{macro2}	
\end{eqnarray}
where
\begin{eqnarray}
&&\hspace{-1cm}  U =  c_1 \Om - \Phi_0 \nabla_x \rho , \quad V =  c_2 \Om - \Phi_0 \nabla_x \rho ,
\label{macro21}\\
&&\hspace{-1cm}  p(\rho ) = d \rho  + \alpha \Phi_0 \big( (n-1)d + c_2 \big) \dfrac{\rho^2}{2}, \quad \gamma = k_0 \big( (n-1)d + c_2 \big).
\label{macro22}
\end{eqnarray}
and the coefficients $c_1, c_2$ will be defined in formulas (\ref{c1}), (\ref{c2}) below.		
\end{theorem}

\noindent
Going back to unscaled variables, we find the model (\ref{macro1_intro})-(\ref{macro5_intro}) presented in the Introduction.

\noindent
{\bf{Proof:} }
 The proof of this theorem is divided into three steps: (i) determination of the equilibrium states ; (ii) determination of the Generalized Collision Invariants ; (iii) hydrodynamic limit. We give a sketch of the proof for each step.

\medskip
\noindent
{\bf Step (i): determination of the equilibrium states} We define the equilibria as the elements of the null space of $Q$, considered as an operator acting on functions of $\omega$ only. 

\begin{definition} \label{def:small_equi}
The set ${\mathcal E}$ of equilibria of $Q$ is defined by
$$
{\mathcal E} = \big\{ f \in H^1({\mathbb S}^{n-1}) \, \, | \, \, f \geq 0 \mbox{ and } Q(f) = 0 \big\} .
$$
\end{definition}
\noindent

We have the following:

\begin{lemma}
The set ${\mathcal E}$ is given by
$$
\mathcal E = \Big\{  \rho M_\Om(\o) \, \, | \, \, \rho \in {\mathbb R}_+, \, \, \Omega \in {\mathbb S}^{n-1} \Big\}
$$
\label{lem:smallkin_equi}
\end{lemma}

\noindent
\vspace{-0.2cm}
For a proof of this lemma, see \cite{Degond_Motsch_M3AS08}. The proof relies on writing the collision operator as
$$
 Q(f) = \nabla_\omega \cdot \Big( M_{\Omega_f} \nabla_\omega \big( \frac{f}{M_{\Omega_f}} \big) \Big). 
$$

\medskip
\noindent
{\bf Step (ii): Generalized Collision Invariants (GCI).} We begin with the definition of a collision invariant. 

\begin{definition}
A collision invariant (CI) is a function $\psi(\o)$ such that for all functions $f(\o)$ with sufficient regularity we have
$$
\int_{\o \in {\mathbb S}^{n-1}} Q (f) \, \psi \, d\o  = 0.
$$
We denote by ${\mathcal C}$ the set of CI. The set ${\mathcal C}$ is a vector space.
\label{def:CI}
\end{definition}

\noindent
As seen in \cite{Degond_Motsch_M3AS08}, the space of CI is one dimensional and spanned by the constants. Physically, this corresponds to conservation of mass during particle interactions. Since energy and momentum are not conserved, we cannot hope for more physical conservations. Thus the set of CI is not large enough to allow us to derive the evolution of the macroscopic quantities $\rho$ and $\Om.$ To overcome this difficulty, a weaker concept of collision invariant, the so-called ``Generalized collisional invariant'' (GCI) has been introduced in \cite{Degond_Motsch_M3AS08}. To introduce this concept, we first define the operator ${\mathcal Q}(\Omega, f)$, which, for a given $\Omega \in {\mathbb S}^{n-1}$, is given by
$$
 {\mathcal Q}(\Omega, f) = \nabla_\omega \cdot \Big( M_\Omega \nabla_\omega \big( \frac{f}{M_\Omega} \big) \Big). 
$$
We note that 
\begin{equation}
Q(f) = {\mathcal Q}(\Omega_f, f), 
\label{eq:calQ_def}
\end{equation}
and that for a given $\Omega \in {\mathbb S}^{n-1}$, the operator $f \mapsto 
 {\mathcal Q}(\Omega, f)$ is a linear operator. Then we have the

\begin{definition}
Let $\Omega \in {\mathbb S}^{n-1}$ be given. A Generalized Collision Invariant (GCI) associated to $\Omega$ is a function $\psi  \in  H^1({\mathbb S}^{n-1})$ which satisfies:
\begin{equation}
\int_{\o \in {\mathbb S}^{n-1}} {\mathcal Q} (\Omega, f) \, \psi(\o) \, d\o  = 0, \quad \forall f \in H^1({\mathbb S}^{n-1}) \quad \mbox{ such that } \quad P_{\Omega^\bot} \Omega_f  = 0. 
\label{eq:sm_def_GCI}
\end{equation}
We denote by ${\mathcal G}_\Omega$ the set of GCI associated to $\Omega$.
\label{def:GCI}
\end{definition}

\noindent
The following Lemma characterizes the set of generalized collision invariants.

\begin{lemma}
There exists a positive function $h$: $[-1,1] \to {\mathbb R}$ such that 
\begin{equation*}
{\mathcal G}_\Omega  =  \{  C +  h(\omega \cdot \Omega) \beta \cdot \omega \, \, \mbox{with arbitrary} \, \,  C\in \R \mbox{ and } \beta \in {\mathbb R}^n \mbox{ such that } \beta \cdot \Omega = 0 \}.
 \end{equation*}
The function $h$ is such that $h(\cos \theta) = \frac{g(\theta)}{ \sin \theta}$ and $ g(\theta)$ is the unique solution in the space $V$ defined by
\begin{eqnarray*}
&&\hspace{-1cm}
V = \{ g \, \, \vert \, \,  (n-2)(\sin \theta)^{\frac{n}{2} - 2} g \in  L^2(0, \pi), \quad (\sin \theta)^{\frac{n}{2} - 1} g \in H^1_0(0, \pi) \},
 \end{eqnarray*}
(denoting by $H^1_0(0, \pi) $ the Sobolev space of functions which are square integrable as well as their derivative and vanish at the boundary) of the problem
\begin{eqnarray*}
&&\hspace{-1cm}
-\sin^{2-n} \theta  \, e^{ -\frac{\cos\theta}{d}} \,  \frac{d}{d \theta} \big(  \sin^{n-2} \theta \, e^{ \frac{\cos\theta}{d}} \, \,\frac{dg}{d \theta}(\theta) \big) + \dfrac{n-2}{\sin^2 \theta } \,  g(\theta)= \sin \theta.
 \end{eqnarray*}
The set ${\mathcal G}_\Omega$ is a $n$-dimensional vector space.
\end{lemma}
\noindent
For a proof we refer to \cite{Degond_Motsch_M3AS08} for $n=3$ and to \cite{Frouvelle_M3AS12} for general $n \geq 2$. We denote by $\psi_\Omega$ the vector GCI 
\begin{equation}
\psi_\Omega =  h(\omega \cdot \Omega) \, P_{\Omega^\bot} \omega, 
\label{eq:psi}
\end{equation}

\noindent
We note that, thanks to (\ref{eq:calQ_def}) and (\ref{eq:sm_def_GCI}), we have
\begin{equation}
\int_{\o \in {\mathbb S}^{n-1}} Q(f) \, \psi_{\Omega_f}(\o) \, d\o  = 0, \quad \forall f \in H^1({\mathbb S}^{n-1}). 
\label{eq:intGCI}
\end{equation}

\medskip
\noindent
{\bf Step (iii): Hydrodynamic limit $\ep \rightarrow 0.$} In the limit $\ep \rightarrow 0$,  we assume that (\ref{eq:converg}) holds. Then , thanks to (\ref{lastscaling1}), we have $Q(f) = 0 $. In view of Lemma \ref{lem:smallkin_equi}, this implies that $f$ has the form (\ref{f0}). We now need to determine the equations satisfied by $\rho$ and $\Omega$. 

For this purpose, we divide Eq. (\ref{lastscaling1}) by $\ep$ and integrate it with respect to $\o$.  Writing (\ref{lastscaling1}) as 
\begin{equation}
({\mathcal T}_1+{\mathcal T}_2 + {\mathcal T}_3)f^{\varepsilon}=\dfrac{1}{\ep} \, Q(f^\ep), 
\label{eq:kin_reduced}
\end{equation}
where 
\begin{equation}
{\mathcal T}_1 f =\partial_t f + \nabla_x \cdot (v_f  f), \quad {\mathcal T}_2 f = \alpha \, \nabla_\o \cdot (P_{\o^\perp} \, v_f \, f), \quad {\mathcal T}_3 f =  \nabla_\o \cdot (P_{\o^\perp} \,  G^1_{f} \, f), 
\label{eq:T123}
\end{equation}
we observe that  the integral of ${\mathcal T}_2 f^\varepsilon$ and ${\mathcal T}_3 f^\varepsilon$over $\o$ is zero since it is in divergence form and the integral of the right-hand side of (\ref{eq:kin_reduced}) is zero since $1$ is a CI. The integral of ${\mathcal T}_1 f^\varepsilon$ gives: 
$$ \partial_t \rho_{f^\varepsilon} + \nabla_x \cdot \Big( \int_{{\mathbb S}^{n-1}} f^\varepsilon(x,\omega,t) \, v_{f^\varepsilon} (x,\omega,t) \, d\omega \Big) = 0. $$
We take the limit $\ep \rightarrow 0$ and use (\ref{eq:converg}) to get Eq. (\ref{macro1}) with 
$$ U = \int_{{\mathbb S}^{n-1}} \rho(x,t) \, M_{\Omega(x,t)} (\omega) \, v_{\rho M_\Omega} (x,\omega,t) \, d\omega. $$
Using (\ref{lastscaling2}), we get $v_{\rho M_\Omega} (x,\omega,t) = \omega - \Phi_0 \nabla_x \rho(x,t)$.  With (\ref{c1}), this leads to the first equation (\ref{macro21}).

Multiplying (\ref{eq:kin_reduced}) by $\psi_{\Om_{f^\ep}}$, integrating with respect to $\o$ and using (\ref{eq:intGCI}), we get
$$ \int_{{\mathbb S}^{n-1}} ({\mathcal T}_1+{\mathcal T}_2 + {\mathcal T}_3)f^{\varepsilon}(x,\omega,t)  \, \psi_{\Om_{f^\ep}} (x,\omega,t) \, d \omega = 0. 
$$
 and taking the limit $\ep \rightarrow 0,$ we get
\begin{equation}
\label{vitesse}
\int_{{\mathbb S}^{n-1}} ( ({\mathcal T}_1+{\mathcal T}_2 + {\mathcal T}_3) (\rho M_{\Omega}))(x,\omega,t)  \, \psi_{\Om(x,t)} (\omega) \, d \omega = 0. 
\end{equation}
This equation describes the evolution of the mean direction $\Om.$ The computations which lead to (\ref{macro2}) are proved in the Appendix \ref{subsec:small_equi}.
The coefficient $c_2$ in (\ref{macro2}) is defined by
\begin{eqnarray}
\label{c2}
& & c_2(d) = \dfrac{\langle \sin^2 \theta \cos \theta \, h \rangle_{M_\Om} }{ \langle \sin^2 \theta \, h \rangle_{M_\Om}}= \dfrac{\int_{0}^{\pi} \sin^n \theta \cos \theta \, M_\Om \,h \,d\theta}{\int_{0}^{\pi} \sin^n \theta  \,M_\Om \,h \,d\theta},
\end{eqnarray}
where for any function $g(\cos \theta),$ we denote $\langle g \rangle$ by
$$
\langle g \rangle_{M_\Om} = \int_{\o \in \S^{n-1}} M_\Om(\o)\,g(\o \cdot \Om) \,d\o = \dfrac{\int_{0}^{\pi} g(\cos \theta) \,e^{\frac{\cos \theta}{d}}\, \sin^{n-2} \theta \,d\theta}{\int_{0}^{\pi} e^{\frac{\cos \theta}{d}} \,\sin^{n-2} \,\theta \,d\theta}.
$$

\medskip
\begin{remark}
The SOHR model (\ref{macro1}), (\ref{macro2}) can be rewritten as follows
\begin{eqnarray*}
&&\hspace{-1cm}  \partial_t \rho +  c_1 \nabla_x \cdot(\rho \Om)   =  \Phi_0 \Delta_x \left(\dfrac{\rho^2}{2}\right), 
\\
&&\hspace{-1cm} \partial_t \Om +   (\bar V \cdot \nabla_x)\Om +  P_{\Om^\perp}\nabla_x h(\rho) =  \gamma P_{\Om^\perp} \Delta_x \Om, 
\end{eqnarray*}
where the vectors $ \bar V$ and the function $h(\rho)$ are defined by
\begin{eqnarray*}
\bar V =  c_2 \Om - (\Phi_0  + 2 \gamma)  \nabla_x \rho,  \quad h'(\rho) = \frac{1}{\rho} \, p'(\rho), 
\end{eqnarray*}
and where the primes denote derivatives with respect to $\rho$. This writing displays this system in the form of coupled nonlinear advection-diffusion equations. 
 \end{remark}

\setcounter{equation}{0}
\section{Numerical discretization of the SOHR model}
\label{sec:num_solution}
In this section, we develop the numerical approximation of the system (\ref{macro1})-(\ref{macro22}) in the two dimensional case. As mentioned above, this system is not conservative because of the geometric constraint $|\Om| = 1.$ Weak solutions of non-conservative systems are not unique because jump relations across discontinuities are not uniquely defined. This indeterminacy cannot be waived by means of an entropy inequality, by contrast to the case of conservative systems. In \cite{Motsch_Navoret_MMS11} the authors address this problem for the SOH model. They show that the model is a zero-relaxation limit of a conservative system where the velocity $\Omega$ is non-constrained (i.e. belongs to ${\mathbb R}^n$). Additionally, they show that the numerical solutions build from the relaxation system are consistent with those of the underlying particle model, while other numerical solutions built directly from the SOH model are not. Here we extend this idea to the SOHR model. More precisely, we introduce the following relaxation model (in dimension $n=2$):
\begin{eqnarray}
&&\hspace{-1cm}   \partial_t \rho^\eta +   \nabla_x \cdot (\rho^\eta U^\eta) = 0, \label{relax1}\\
&&\hspace{-1cm}    \partial_t(\rho^\eta \Om^\eta) +  \nabla_x \cdot  ( \rho^\eta V^\eta \otimes \Om^\eta)  +     \nabla_x p( \rho^\eta )  - \gamma \Delta_x(\rho^\eta \Om^\eta) = \dfrac{\rho^\eta}{\eta} (1 - |\Om^\eta|^2) \Om^\eta,  \label{relax2} \\
&&\hspace{-1cm}   U^\eta =  c_1 \Om^\eta - \Phi_0 \nabla_x \rho^\eta , \quad V^\eta =  c_2 \Om^\eta - \Phi_0 \nabla_x \rho^\eta,\label{relax22}\\
&&\hspace{-1cm}   p(\rho^\eta) =  d \rho^\eta  + \alpha \Phi_0 \big( d + c_2 \big) \dfrac{(\rho^\eta)^2}{2}, \quad \gamma = k_0 \big( d + c_2 \big).
\label{relax23}
\end{eqnarray}
The left-hand sides form a conservative system. We get the following proposition:

\begin{proposition}
\label{prop1}
The relaxation model (\ref{relax1})-(\ref{relax23}) converges to the SOHR model (\ref{macro1})-(\ref{macro22}) as $\eta $ goes to zero.
\end{proposition}

\noindent
The proof of proposition \ref{prop1} is given in Appendix \ref{numerical:convergence}. This allows us to use well-established numerical techniques for solving the conservative system (i.e. the left-hand side of (\ref{relax1}), (\ref{relax2})). The scheme we propose relies on a time splitting of step $\Delta t$ between the conservative part
\begin{eqnarray}
&&\hspace{-1cm}   \partial_t \rho^{\eta} +   \nabla_x \cdot (\rho^{\eta} U^{\eta}) = 0, 
\label{split1}\\
&&\hspace{-1cm}    \partial_t(\rho^{\eta} \Om^{\eta}) +  \nabla_x \cdot  ( \rho^{\eta} V^{\eta} \otimes \Om)  +     \nabla_x  p(\rho^{\eta})   - \gamma \Delta_x (\rho^{\eta} \Om^{\eta}) = 0, 
\label{split2}
\end{eqnarray}
and the relaxation part
\begin{eqnarray}
&&\hspace{-1cm}\partial_t \rho^\eta = 0, 
\label{relax3} \\
&&\hspace{-1cm}\partial_t(\rho^\eta \Om^\eta)  = \dfrac{\rho^\eta}{\eta} (1 - |\Om^\eta|^2) \Om^\eta. 
\label{relax4}
\end{eqnarray}

\medskip
\noindent
System (\ref{split1}-\ref{split2}) can be rewritten in the following form (we omit the superscript $\eta$ for simplicity)
$$
	Q_t + (F(Q, Q_x))_x + (G(Q, Q_y))_y = 0,
$$
where
$$
	Q =  \begin{pmatrix}
				\rho \\ \rho \Om_1 \\ \rho \Om_2
		\end{pmatrix}, \quad
	F(Q, Q_x) =  \begin{pmatrix}
				 \rho U_1  \\
				\rho \Om_1 V_1 + p(\rho) - \gamma \partial_x(\rho \Om_1),  \\
				\rho \Om_1 V_2 - \gamma \partial_x(\rho \Om_2)
		\end{pmatrix}, 
$$
$$
	G(Q, Q_y) =  \begin{pmatrix}
				 \rho U_2  \\
				 \rho \Om_2 V_1 - \gamma \partial_y(\rho \Om_1) \\
				\rho \Om_2 V_2 + p(\rho) - \gamma \partial_y(\rho \Om_2)
				\end{pmatrix}.
$$

\medskip
\noindent
We consider now the following numerical scheme where we denoted $Q^*_{i,j}$ the approximation of $Q$ at time $t^{n+1} = (n+1) \Delta t $ and position $x_i = i \Delta x, y_j = j \Delta y$:
\begin{eqnarray*}
 Q^{*}_{i,j} & = & Q^{n}_{i, j} - \dfrac{\Delta t}{\Delta x} \{F^n_{i +1/2,j} - F^n_{i-1/2,j}\}  
- \dfrac{\Delta t}{\Delta y} \{ G^n_{i,j+1/2} - G^n_{i,j-1/2} \}, 
\end{eqnarray*}
where the numerical flux $F^n_{i +1/2,j}$ is given by
\begin{equation*}
	 F^n_{i +1/2,j}  = \dfrac{F^n(Q^n_{i,j}, Q^n_{xi, j}) + F^n(Q^n_{i+1,j}, Q^n_{x(i+1),j})}{2} - P^{i+\frac{1}{2}}_2 \Big( \dfrac{\partial F}{\partial Q}(\bar Q^n_{i,j}, \bar{Q}^n_{xi,j} ) \Big) (Q^n_{i+1,j} - Q^n_{i,j}),
\end{equation*}
with
\begin{equation*}
Q^n_{xi,j}  =  \frac{(Q^n_{i+1,j} - Q^n_{i,j} )}{\Delta x} , \quad \bar{Q}^n_{i,j} = \dfrac{Q^n_{i,j} + Q^n_{i+1, j}}{2},	 \quad \bar{Q}^n_{xi,j} = \dfrac{Q^n_{xi,j} + Q^n_{x(i+1), j}}{2},
\end{equation*}
and the analogous discretization holds for $G^n_{i ,j+ \frac{1}{2}}$.

\noindent
In the above formula, $P^{i+\frac{1}{2}}_2$ is a polynomial of matrices of degree $2$ calculated with the eigenvalues of the Jacobian matrices $\dfrac{\partial F}{\partial Q}$ at an intermediate state depending on $(Q^n_{i,j}, Q^n_{xi, j})$ and $(Q^n_{i+1,j}, Q^n_{x(i+1), j})$ as detailed in \cite{Degond_Peyrard_99}.
To ensure stability of the scheme, the time step $\Delta t$ satisfies a Courant-Friedrichs-Lewy (CFL) condition computed as the minimum of the CFL conditions required for the hyperbolic and diffusive parts of the system.

Once the approximate solution of the conservative system is computed, equations (\ref{relax3}) and (\ref{relax4}) can be solved explicitly. In the limit $\eta\rightarrow 0$ they give
$$
    \rho^{n+1}=\rho^{*}, \qquad \Om^{n+1}=\frac{\Om^{*}}{|\Om^{*}|}
$$
where $(\rho^*,\Om^{*})$ is the numerical solution of system (\ref{split1}-\ref{split2}). This ends one step of the numerical scheme for the system (\ref{relax1}-\ref{relax2}).

\section{Numerical tests}
The goal of this section is to present some numerical solutions of the system (\ref{macro1})-(\ref{macro22}) which validate the numerical scheme proposed in the previous section. We will first perform a convergence test. We then successively compare the solutions obtained with the SOHR model with those computed by numerically solving the individual based model (\ref{overdamped1}) in regimes in which the two models should provide similar results. We will finally perform some comparisons between the SOH and the SOHR system to highlight the difference between the two models. We will compare the SOHR model with another way to incorporate repulsion in the SOH Model, the so-called DLMP model of \cite{Degond_etal_MAA13}.

For all the tests, we use the model in uscaled variables as described in the Introduction (see (\ref{macro1_intro})-(\ref{macro5_intro}). The potential kernel $\phi$ is chosen as
\begin{equation}
\label{phi}
\phi(x) = \begin{cases}
 (|x| -1)^2  \quad \mbox{if} \quad |x| \leq1,\\
0 \quad \mbox{if} \quad  |x| > 1,
\end{cases}
\end{equation}
which gives $\Phi_0 = \dfrac{\pi}{6}$, while for $K$, by assumption normalized to $1,$ we choose the following form
$$
K(\vert z \vert ) = \begin{cases}
 \dfrac{1}{\pi}  \quad \mbox{if} \quad |z| \leq 1,\\
0 \quad \mbox{if} \quad  |z| > 1.
\end{cases}
$$
This leads to $k_0 = \dfrac{1}{8}.$ The other parameters, which are fixed for all simulations if not differently stated, are :
$$ v_0 = 1, \,\, \mu = \dfrac{1}{2}, \,\, \alpha = 1, \,\, d = 0.1, \,\, L_x = 10, \,\,  L_y = 10, $$
which, in dimension $n=2$, lead to (after numerically computing the GCI and the associated integrals):
$$ c_1 = 0.9486, \,\,  c_2 =0.8486. $$
In the visualization of the results, we will use the angle $\theta$ of the vector $\Omega$ relative to the $x$-axis, i.e. $\Omega = (\cos \theta, \sin \theta)$. 

\subsection{Convergence test}
\begin{figure}[!ht]
\centering
\includegraphics[height=8cm,width=11cm]{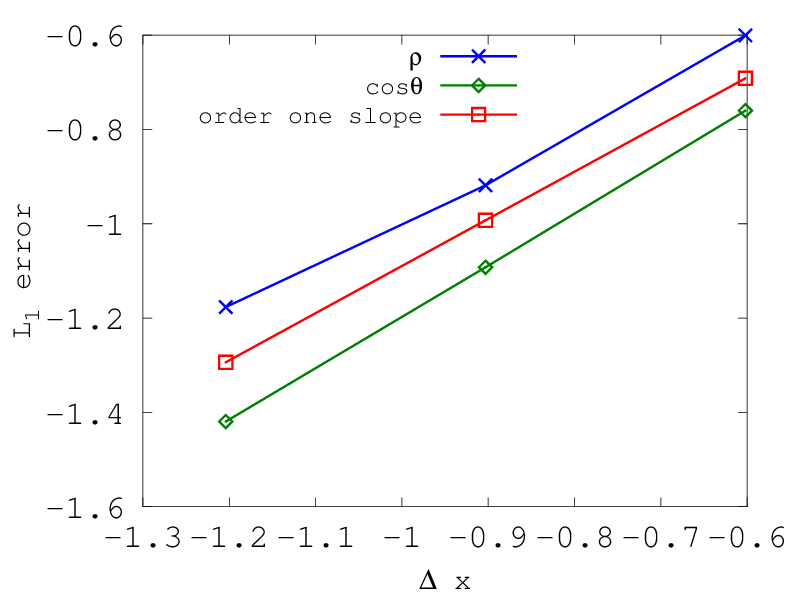}
\caption{$L^1$-error for the density $\rho$ and the flux direction $ \cos\theta $ as a function of $\Delta x$ in log-log scale. A straight line of slope $1$ is plotted for reference. This figure shows that the scheme is numerically of order $1$. 
}
\label{fig:scheme}
\end{figure}

The first test is targeted at the validation of the proposed numerical scheme. For this purpose, we investigate the convergence when the space step $(\Delta x, \Delta y )$ tends to $(0,0)$, refining the grid and checking how the error behaves asymptotically. The initial mesh size is $\Delta x = \Delta y = 0.25$ while the time step is $\Delta t = 0.001.$ We repeat the computation for $(\dfrac{\Delta x}{2}, \dfrac{\Delta y}{2}),$ $(\dfrac{\Delta x}{4}, \dfrac{\Delta y}{4}),$ $(\dfrac{\Delta x}{8}, \dfrac{\Delta y}{8})$. The convergence rate is estimated through the measure of the $L^1$ norm of the error for the vectors $(\rho, \cos\theta)$ by using for each grid the next finer grid as reference solution. The initial data is
\begin{equation}
\label{Vortex}
\rho_0 = 1, \quad  \theta_0(x,y) = \begin{cases}
										\arctan(\dfrac{y_1}{x_1} ) + \dfrac{\pi}{2} sign(x_1) \,\, \mbox{if} \,\, x_1 \neq 0, \\
										\pi \,\,\mbox{if} \,\, x_1 = 0 \,\, \mbox{and} \,\, y_1 > 0, \\
                                         0  \,\,  \mbox{if} \,\,  x_1 = 0  \,\, \mbox{and} \,\,  y_1 < 0,
                                   \end{cases}
\end{equation}
where
$$ x_1  =   x - \dfrac{L_x}{2}, \quad  y_1 =   y - \dfrac{L_y}{2}.$$
The boundary conditions are fixed in time on the four sides of the square : $( \rho^n, \theta^n ) = (\rho_0, \theta_0 )$.
The error curves for the density and for $ \cos \theta $ are plotted in figure \ref{fig:scheme} as a function of the space step in log-log scale at time $T = 1s.$  The slope of the error curves are compared to a straight line of slope 1.
From the figure, we observe the convergence of the scheme with accuracy close to $1.$

\noindent

\subsection{Comparison between the SOHR and the Vicsek model with repulsion}

In this subsection, we validate the SOHR model by comparing it to the Vicsek model with repulsion on two different test cases. We investigate the convergence of the microscopic IBM to the macroscopic SOHR model when the scaling parameter $\ep$ tend to zero. The scaled IBM is written: 
\begin{eqnarray*} 
&&\hspace{-1cm} \dfrac{dX_k}{dt} = v_k, \qquad 
v_k	=  \o_k  -  \, \nabla_x \Phi(X_k(t),t),\\
&&\hspace{-1cm} d\o_k	=   P_{\o_k^\perp} \circ \Big( \frac{1}{\varepsilon} \, \bar{\o}(X_k(t),t) \, dt  + \alpha \, v_k \, dt + \sqrt{\frac{2d}{\varepsilon}} \, dB^k_t \Big).\\
&&\hspace{-1cm} \Phi(x,t) = \dfrac{1}{\varepsilon^2 \, N} \sum_{i=1}^{N} \nabla \phi \Big( \frac{|x - X_i|}{\varepsilon \, r} \Big), \\
&&\hspace{-1cm} \bar{\o}(x,t) =  \dfrac{{\mathcal J}(x,t)}{\vert {\mathcal J}(x,t) \vert},  \qquad   {\mathcal J} (x,t)  = \frac{1}{N} \sum\limits_{i=1}^{N} K\Big(\frac{|x-X_i|}{\sqrt{\varepsilon} \, R}\Big) \o_i. 
\end{eqnarray*}	
The solution of the individual based model (\ref{overdamped1}-\ref{overdamped3}) is computed by averaging different realizations in order to reduce the statistical errors. The coefficient of the IBM are fixed to $r = 0.0625$ for the repulsive range, $R = 0.25$ for the alignment interaction range, while $N = 10^5$ particles are used for each simulation. The details of the particles simulation can be found in \cite{Fehske_2007, Hockney_1988} for classical particle approaches or in \cite{Motsch_Navoret_MMS11} for a direct application to the SOH model.

\begin{figure}[!ht]
\subfloat[For density $\rho$ \label{dd}]{%
\includegraphics[width=0.51\textwidth]{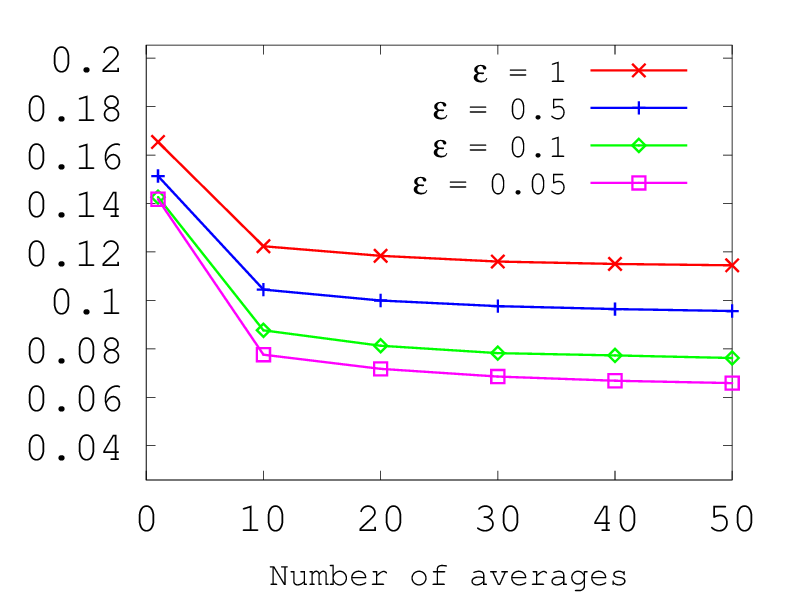}
    }
\hfill
\subfloat[For $\theta$  \label{velo}]{%
\includegraphics[width=0.51\textwidth]{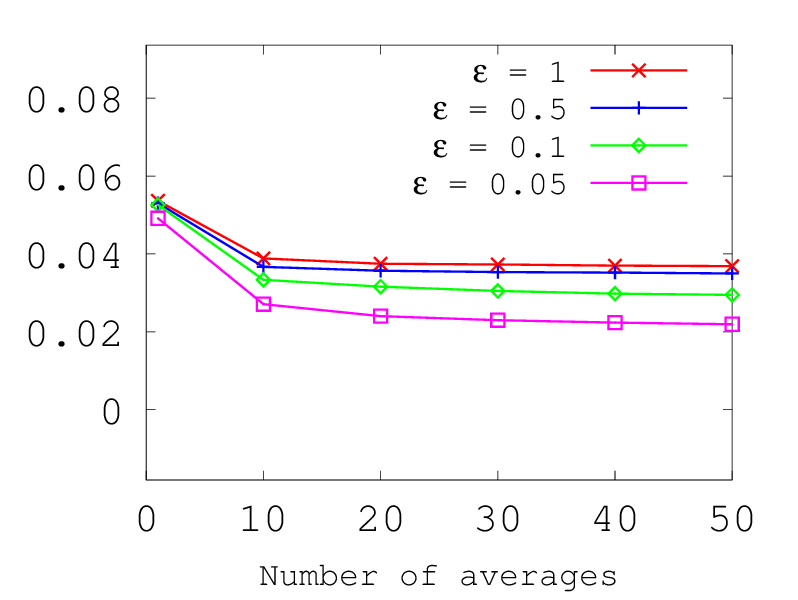}
    }
\caption{Relative error between the macroscopic and the microscopic model for density (a) and $ \theta $ (b) as a function of the number of averages for different values of $\ep$. The error decreases with both decreasing $\varepsilon$ and increasing number of averages, showing that the SOHR model provides a valid approximation of the IBM for $\rho$ and $\theta$.  }
\label{fig:macro_micro}
\end{figure}

\begin{figure}[!ht]
\subfloat[density $\rho$ \label{density}]{%
\includegraphics[width=0.5\textwidth]{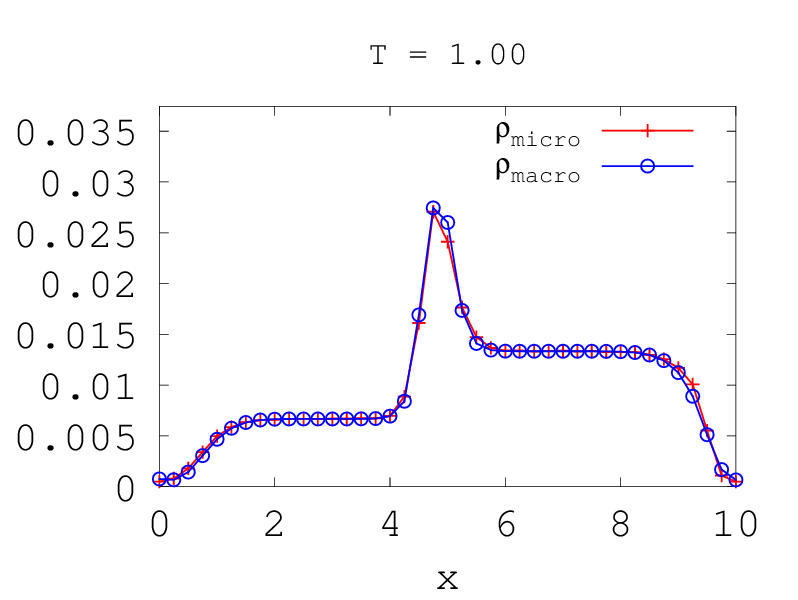}
    }
\hfill
\subfloat[ $\theta$  \label{velocity}]{%
\includegraphics[width=0.5\textwidth]{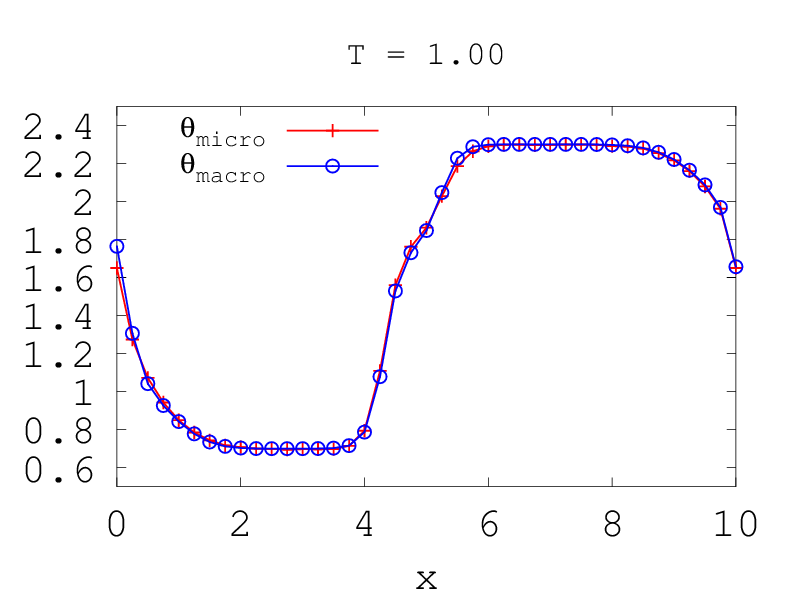}
    }
\caption{Solution of the Riemann problem (\ref{RM}) along the $x$-axis for the SOHR model (blue line) and for the IBM with $\varepsilon = 0.05$ (red line) at $T = 1s.$ The agreement between the two models is excellent. For the SOHR model, the mesh size is $\Delta x = \Delta y = 0.0625$. }
\label{fig:solution_RM}
\end{figure}

\medskip
\noindent
\paragraph{Riemann problem:}
The convergence of the two models is measured on a Riemann problem with the following initial data
\begin{equation}
\label{RM}
(\rho_l, \theta_l ) = (0.0067, 0.7 ) ,\quad  (\rho_r, \theta_r ) = (0.0133, 2.3).
\end{equation}
and with periodic boundary condition in $x$ and $y$. The parameters of the SOHR model are: $\Delta t = 0.01, \,\, \Delta x = \Delta y = 0.25$. In figure \ref{fig:macro_micro} we report the relative $L^1$ norm of the error for the macroscopic quantities $(\rho, \theta)$  between the SOHR model and the particle model with respect to the number of averages for different values of $\ep$ : $ \ep = 1$ (x-mark), $\ep = 0.5$ (plus), $\ep = 0.1$ (circle), $\ep = 0.05$ (square) at time $T = 1s.$ This figure shows, as expected, that the distance between the two solutions goes to zero when $\ep$ goes to zero.
In figure \ref{fig:solution_RM} we report the density $\rho$ and the flux direction $\theta$ for the same Riemann problem along the $x$-axis for $\ep = 0.05$ at time $T = 1s$, the solution being constant in the $y$-direction. Again we clearly observe that the two models provide very close solutions, the small differences being due to the different numerical schemes employed for their discretizations.

\begin{figure}[!ht]
\subfloat[ density $\rho$ for the IBM \label{density_micro_t06}]{%
\includegraphics[width=0.5\textwidth]{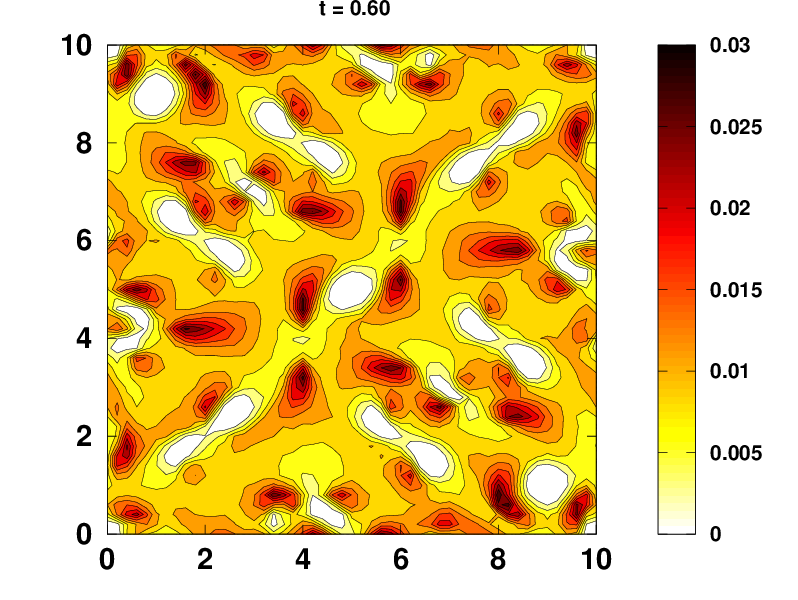}
    }
\hfill
\subfloat[density $\rho$ for the SOHR model  \label{density_macro_t06}]{%
\includegraphics[width=0.5\textwidth]{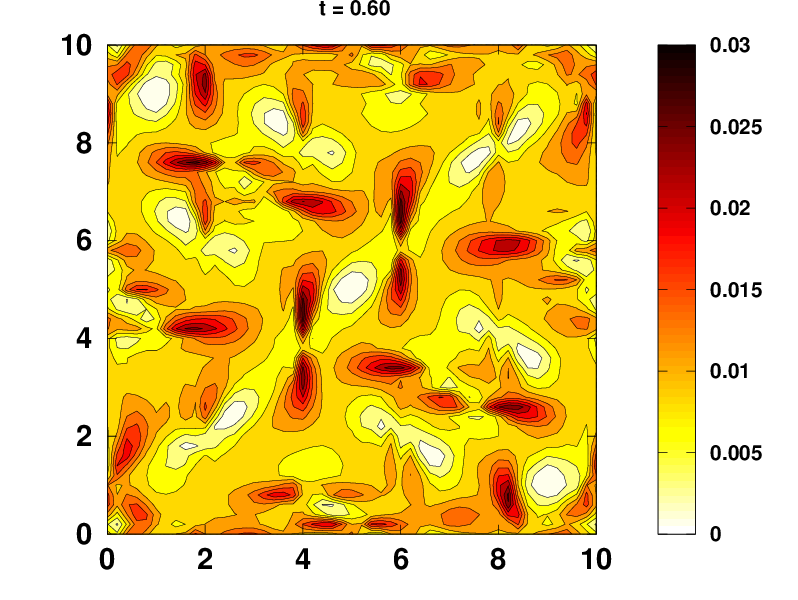}
    }
\caption{Density $\rho$ for Taylor-Green vortex problem \ref{Taylor_Green} at time $t = 0.6s$. Left: IBM. Right: SOHR model.  }
\label{fig:dens_macro_micro_t06_GT}
\end{figure}

\begin{figure}[!ht]
\subfloat[ $\Om$ for the IBM \label{velo_micro_t06}]{%
\includegraphics[width=0.51\textwidth]{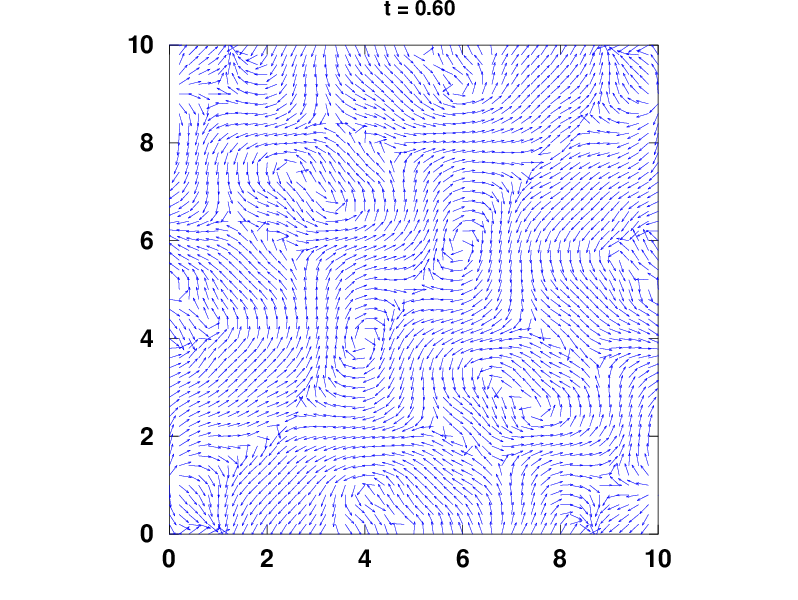}
    }
\hfill
\subfloat[$\Om$ for the SOHR model  \label{velo_macro_t06}]{%
\includegraphics[width=0.51\textwidth]{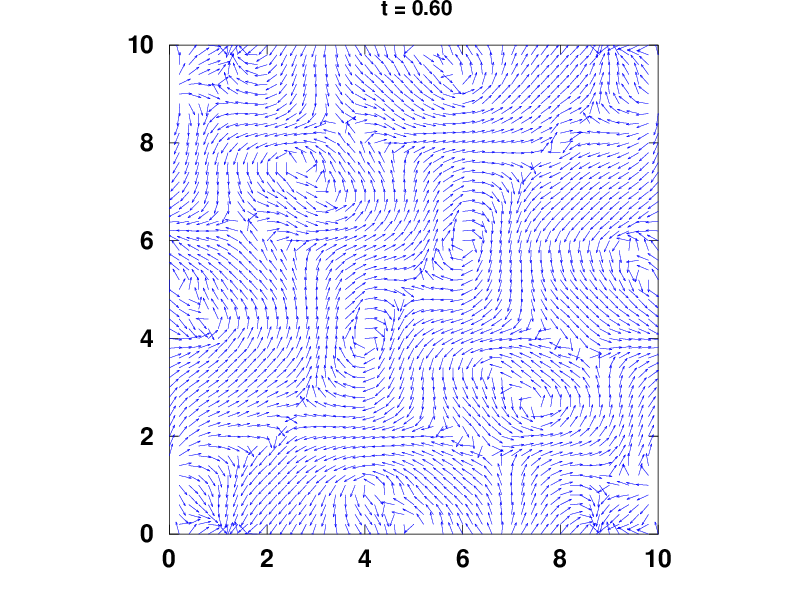}
    }
\caption{Mean direction $\Om$ for Taylor-Green vortex problem \ref{Taylor_Green} at time $t = 0.6s$. Left: IBM. Right: SOHR model. }
\label{fig:velo_macro_micro_t06_GT}
\end{figure}

\medskip
\noindent
\paragraph{Taylor-Green vortex problem:}
In this third test case, we compare the numerical solutions provided by the two models in a more complex case. The initial data are
\begin{equation}
\label{Taylor_Green}
\rho_0 = 0.01, \quad  \Om_0(x,y) = \dfrac{\tilde{\Om}_0(x,y)}{\vert \tilde{\Om}_0(x,y) \vert},
\end{equation}
 where the vector $ \tilde{\Om}_0 = (\tilde{\Om}_{01}, \tilde{\Om}_{02}) $ is given by
\begin{eqnarray*}		
	&& \tilde{\Om}_{01}(x,y) = \dfrac{1}{3}\sin(  \dfrac{\pi}{5} x ) \cos(  \dfrac{\pi}{5} y ) + \dfrac{1}{3} \sin(  \dfrac{3\pi}{10} x) \cos( \dfrac{3\pi}{10} y ) + \dfrac{1}{3} \sin(  \dfrac{\pi}{2} x) \cos( \dfrac{\pi}{2} y),\\
	 && \tilde{\Om}_{02}(x,y) = - \dfrac{1}{3} \cos( \dfrac{\pi}{5} x) \sin(  \dfrac{\pi}{5} y )- \dfrac{1}{3} \cos( \dfrac{3\pi}{10} x ) \sin( \dfrac{3 \pi}{10} y) - \dfrac{1}{3} \cos( \dfrac{\pi}{2} x) \sin(  \dfrac{\pi}{2} y ).	
\end{eqnarray*}
with periodic boundary conditions in both directions. The numerical parameters for the SORH model are : $\Delta x = \Delta y  = 0.2,$ $\Delta t = 0.01$, while for the particle simulations we choose : $N = 10^5$ particles, $\ep = 0.05,$  $r = 0.04,$ $R = 0.2.$  
In figure \ref{fig:dens_macro_micro_t06_GT} and \ref{fig:velo_macro_micro_t06_GT}, we report the density $\rho$ and the flux direction $\Om$ at time $t = 0.6$s. In both figures, the left picture is for the IBM and the right one for the SOHR model. Again, we find a very good agreement between the two models in spite of the quite complex structure of the solution.


\subsection{Comparison between the SOH and the SOHR model}
In this part, we show the difference between the SOH system (\ref{macro6_intro}), (\ref{macro7_intro}) and the SOHR one for different values of the repulsive force $\Phi_0.$ We recall that the SOHR model reduces to the SOH one in the case in which the repulsive force is set equal to zero. To this aim, we rescale the repulsive force $\Phi_0$ by
$$ \Phi_0 = F_0 \int_{x \in \R^2} \phi(x) dx $$
and then we let $F_0$ vary. The repulsive potential $\phi$ is still given by (\ref{phi}), so that $\Phi_0 = F_0 \pi/6$.
The other numerical parameters are chosen as follows:
$ d = 0.05,$ $\alpha = 0,$ $ k_0 = 1/8, $  $\mu = 1,$ $  L_x = 10, L_y = 10,$ $  \Delta x = \Delta y = 0.15,$ $ \Delta t = 0.001.$
The initial data are those of the vortex problem (\ref{Vortex}) except that we start with four vortices instead of only one. Periodic boundary conditions in both directions are used.

\begin{figure}[!ht]
\subfloat[density $\rho$ for $F_0 = 5$ \label{F5_density}]{%
\includegraphics[width=0.51\textwidth]{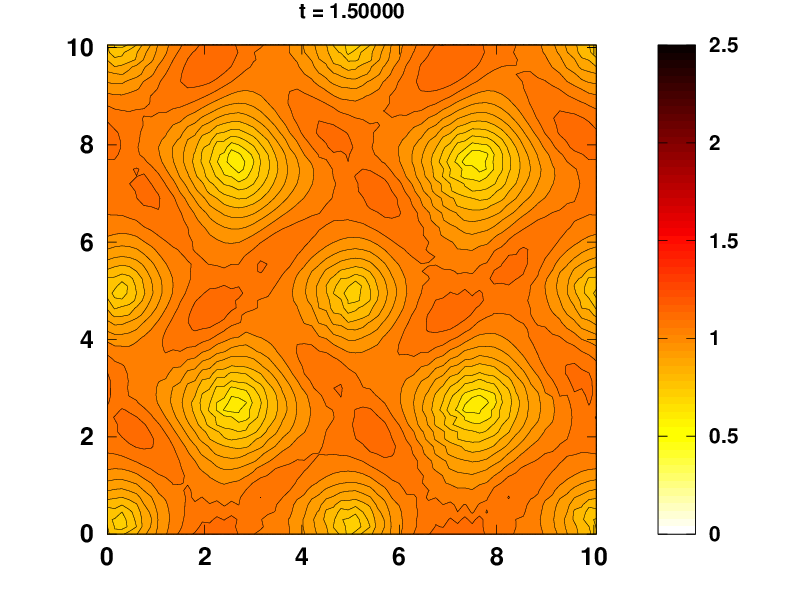}
    }
\hfill
\subfloat[$\Om$ for $F_0 = 5$  \label{F5_velo}]{%
\includegraphics[width=0.51\textwidth]{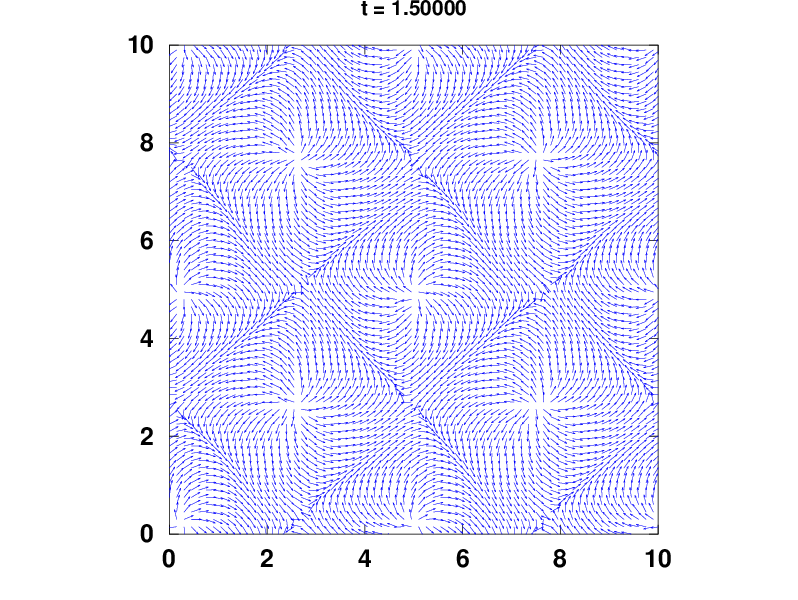}
    }
\caption{Solution of the SOHR model for $F_0 = 5.$ Density $\rho$ (fig \ref{F5_density} ), flux direction $\Om$ (fig.\ref{F5_velo} ) at $t= 1.5s$.}
\label{fig:solution_F5}
\end{figure}

\begin{figure}[!ht]
\subfloat[ density $\rho$ for $F_0 = 0.05$ \label{F005_density}]{%
\includegraphics[width=0.51\textwidth]{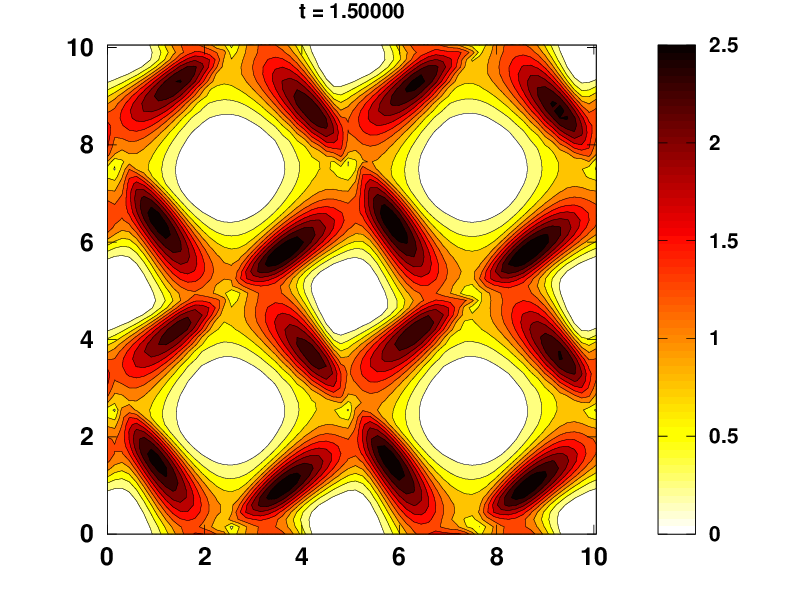}
    }
\hfill
\subfloat[ $\Om$ for $F_0 = 0.05$ \label{F005_velo}]{%
\includegraphics[width=0.51\textwidth]{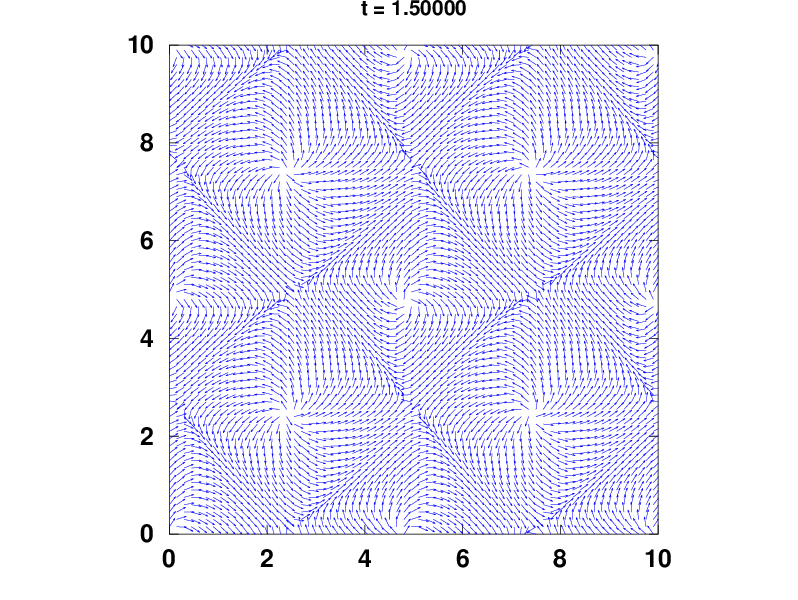}
}
\caption{Solution of the SOHR model for $F_0 = 0.05.$ Density $\rho$ (fig \ref{F5_density} ), flux direction $\Om$ (fig.\ref{F5_velo} ) at $t = 1.5s$.}
\label{fig:solution_F005}
\end{figure}

Figure (\ref{fig:solution_F5}) displays the solutions for the SOHR system for the density (left) and for the flux direction
(right) at $T = 1.5s$ with $F_0 = 5.$ Figure (\ref{fig:solution_F005}) displays the solutions for $F_0 = 0.05.$ The results are almost undistinguishable to those of the SOH model ($F_0=0$) and are not shown for this reason. 
These figures show that when the repulsive force is large enough, the SOHR model can prevent the formation of high concentrations. By contrast, when this force is small, the SOHR model becomes closer to the SOH one and high concentrations become possible.

\subsection{Comparison between the SOHR and the DLMP model}

In this final part, we want to compare the SOHR system to the hydrodynamic model proposed by Degond, Liu, Motsch and Panferov in \cite{Degond_etal_MAA13} (referred to as DLMP model). This model is derived, in a similar fashion as the SOHR model, starting from a system of self-propelled particles which obey to alignment and repulsion. The main difference is that in the DLMP model, the particle velocity is exactly equal to the self propulsion velocity but the particles adjust their orientation to respond to repulsion as well as alignment. The resulting model is of SOH type and is therefore written (\ref{macro6_intro}),  (\ref{macro7_intro}), but with an increased coefficient in front of the pressure term $P_{\Omega^\bot} \nabla_x \rho$, this coefficient being equal to $v_0 \, d (1 + \frac{d+c_2}{c_1}F_0)$. The initial conditions and numerical parameters are the same as in previous test

\begin{figure}[!ht]
\subfloat[density $\rho$ \label{F5DLMP_density}]{%
\includegraphics[width=0.51\textwidth]{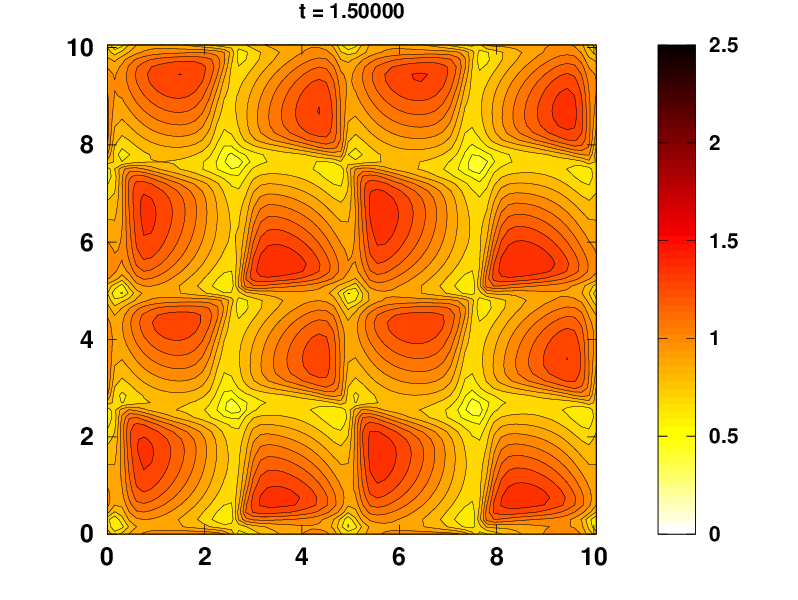}
    }
\hfill
\subfloat[$\Om$  \label{F5DLMP_velo}]{%
\includegraphics[width=0.51\textwidth]{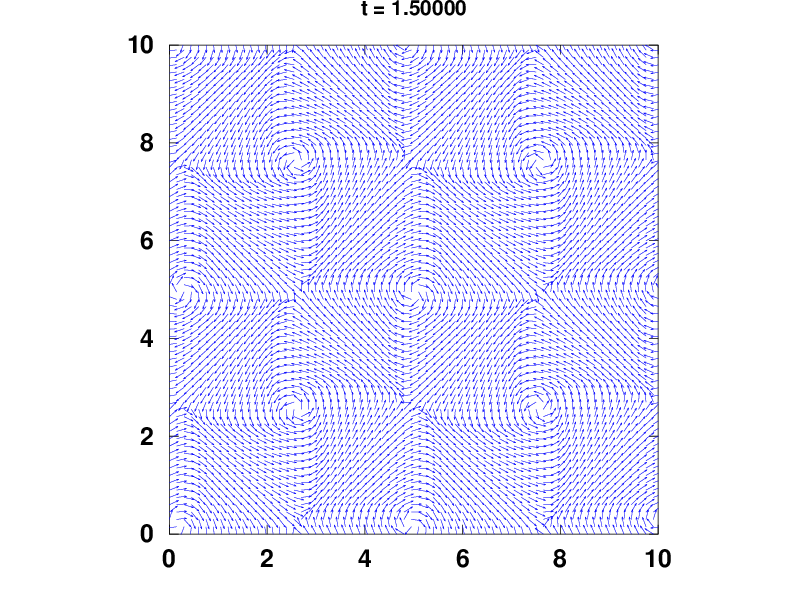}
    }
\caption{Solution of the DLMP model for $F_0 = 5.$ Density $\rho$ (left) and flux direction $\Om$ (right) at $t = 1.5s$.}
\label{fig:solution_DLMP_F5}
\end{figure}

In figure \ref{fig:solution_DLMP_F5}, we report the density $\rho$ (left) and the flux direction $\Om$  (right) for $F_0 = 5$ for the DLMP model. Comparing figures (\ref{fig:solution_F5}) with figure \ref{fig:solution_DLMP_F5}, we observe that the solutions of the SOHR and of the DLMP model are different. The homogenization of the density seems more efficient with the SOHR model than with the DLMP model. This can be attributed to the effect of the nonlinear diffusion terms that are included in the SOHR model but not in the DLMP model. Therefore, the way repulsion is included in the models may significantly affect the qualitative behavior of the solution. In practical situations, when the exact nature of the interactions is unknown, some care must be taken to choose the right repulsion mechanism.

\setcounter{equation}{0}
\section{Conclusion}
\label{sec:conclu}
In this paper, we have derived a hydrodynamic model for a system of self-propelled particles which interact through both alignment and repulsion. In the underlying particle model, the actual particle velocity may be different from the self-propulsion velocity as a result of repulsion interactions with the neighbors. Particles update the orientation of their self-propulsion seeking to locally align with their neighbors as in Vicsek alignment dynamics. The corresponding hydrodynamic model is similar to the Self-Organized Hydrodynamic (SOH) system derived from the Vicsek particle model but it contains several additional terms arising from repulsion. These new terms consist principally of gradients of linear or nonlinear functions of the density including a non-linear diffusion similar to porous medium diffusion. This new Self-Organized Hydrodynamic system with Repulsion (SOHR) has been numerically validated by comparisons with the particle model. It appears more efficient to prevent high density concentrations than other approaches based on simply enhancing the pressure force in the SOH model. In future work, this model will be used to explore self-organized motion in collective dynamics. To this effect, it will be calibrated on data based on biological experiments, such as recordings of collective sperm-cell motion.



\begin{appendices}
\setcounter{equation}{0}
\section{Proof of formulas (\ref{expansion1}), (\ref{expansion2})}
\label{sec:taylorexpasion}
By introducing the change of variable $ z = -\dfrac{x-y}{\sqrt{\ep} R}$ and using Taylor expansion, we get
\begin{eqnarray*}
&  & \hspace{-1cm} \frac{1}{(\sqrt{\ep} R)^n}\int_{\S^{n-1} \times \R^n} K\big(\dfrac{|x - y|}{\sqrt{\ep} R }\big) \, f^\ep(y, \o, t) \, \o \,d\o \,dy  \\
&  & \hspace{-1cm}   =  \int_{\S^{n-1} \times \R^n} K(|z|) \, f^\ep(x + \sqrt{\ep} R  z , \o,t)\,\o\, dz \,d\o \\
&  & \hspace{-1cm}   =   \int_{\S^{n-1} \times \R^n} K(|z|) \, \big(f^\ep + \sqrt{\ep} \,R \,\nabla_x f^\ep \cdot z + \dfrac{\ep R^{2}}{2} D_x^2 f :(z \otimes z) + O(\sqrt{\ep}^3) \big)(x,\o, t) \,\o\, dz\, d\o \\
&  & \hspace{-1cm}  =   \big( J(x,t) + \ep \, k_0 \, \Delta_x J(x,t) + O(\ep^2) \big),
\end{eqnarray*}
where $k_0$ is given by (\ref{k0}) and $D_x^2 f$ is the Hessian matrix of $f$ with respect to the variable $x$. Here, we have used that the ${\mathcal O}(\sqrt{\varepsilon})$ and ${\mathcal O}(\varepsilon^{3/2})$ terms vanish after integration in $z$ by oddness with respect to $z$. 

\medskip
\noindent
By the same computation for the kernel $\phi$, we have
\begin{eqnarray*}
	& &  \frac{1}{(\ep r)^n} \int_{\S^{n-1} \times \R^n}  \phi(\dfrac{|x - y|}{\ep r}) f^\ep(y, \o, t) dy d\o  \\
	 &  & \hspace{1cm} =    \int_{\S^{n-1} \times \R^n} \phi(|z|) f^\ep(x + \ep r z , \o, t) dz d\o \\
	 &  & \hspace{1cm}  =  \int_{\S^{n-1} \times \R^n}  \phi(|z|) (f^\ep + \ep r \nabla_x f \cdot z + O(\ep^2) )(x, \o,t) dz d\o \\
	 &  & \hspace{1cm}  = \Phi_0 \int_{\S^{n-1}}  f^\ep(x, \o,t) d\o + O(\ep^2),
\end{eqnarray*}
with $\Phi_0 =  \int_{\R^n} \phi(|z|) dz. $ \endproof

\section{Proof of Theorem \ref{thm:sm}  }
\label{subsec:small_equi}
%
%
We prove that (\ref{vitesse}) leads to (\ref{macro2}). 
%
%
%
%
Thanks to (\ref{eq:psi}), Eq. (\ref{vitesse}) can be written: 
\begin{equation}
\label{eq2}
P_{\Om^\perp}\int_{\o \in \S^2} ({\mathcal T}_1(\rho M_\Om) + {\mathcal T}_2(\rho M_\Om) +  {\mathcal T}_3 (\rho M_\Om) ) h(\o \cdot \Om) \o d\o :=  T_1 + T_2 + T_3 = 0,
\end{equation}	
%
%
where ${\mathcal T}^k$, $k = 1, 2, 3$ are given by (\ref{eq:T123}). 
%
%
%
%
Now, ${\mathcal T}_1(\rho M_\Om ) $ can be written:
\begin{equation}
\label{t1}
{\mathcal T}_1 (\rho M_\Om ) =  \partial_t(\rho M_\Om)  + \nabla_x \cdot (\o \rho M_\Om) - \Phi_0 \nabla_x \cdot \left(\nabla_x \left(\frac{\rho^2}{2}\right) M_\Om\right).
\end{equation}
We recall that the first two terms of ${\mathcal T_1}$ at the right hand side of (\ref{t1}) and the corresponding terms in $T_1$ have been computed in \cite{Degond_Motsch_M3AS08}. The computation for the third term of $\mathcal T_1 $ is easy and we get:
$$
T_1 =   \beta_1\rho \partial_t \Om  + \beta_2 \rho ( \Om \cdot \nabla_x) \Om + \beta_3 P_{\Om^\perp} \nabla_x \rho  + \beta_4  (\nabla_x \left(\dfrac{\rho^2}{2}\right) \cdot \nabla_x ) \Om
$$
where the coefficients are given by
\begin{eqnarray*}
& & \beta_1 =  \dfrac{1}{d(n-1)} \langle  \sin^2 \theta \, h \rangle_{M_\Om}, \quad  \beta_2 = \dfrac{1}{d(n-1)}\langle  \sin^2 \theta \, \cos \theta \, h \rangle_{M_\Om},   \\
& & \beta_3 = \dfrac{1}{n-1}  \langle  \sin^2 \theta \, h \rangle_{M_\Om}, \quad \beta_4 = -  \dfrac{\Phi_0}{d(n-1)}  \langle  \sin^2 \theta\,  h \rangle_{M_\Om}.
\end{eqnarray*}
Now observe that for a constant vector $A \in \R^n$, we have
\begin{equation}
 \nabla_\o( \o\cdot A) =  P_{\omega^\bot} A,  \qquad  \nabla_\o \cdot  (P_{\omega^\bot} A)  = -(n-1) \o \cdot A  .
\label{eq:nablaomA}
\end{equation}
Thus, using (\ref{eq:T123}), (\ref{eq:nablaomA}) and the chain rule, we get for $\mathcal T_2(\rho M_\Om)$
\begin{eqnarray*}
\mathcal T_2(\rho M_\Om) &  = &  \alpha \Phi_0 \bigg( (n-1) \, \o \cdot \nabla_x\left(\rho^2/2\right)
 - d^{-1}  \nabla_x\left(\rho^2/2\right) \cdot \Om  \\
 &  & \hspace{4cm}  + d^{-1} \big(\o \cdot \nabla_x (\rho^2/2 ) \big) (\o \cdot \Om)  \bigg) M_\Om
\end{eqnarray*}
%
%
%
Finally, we obtain:
$$
T_2 = \beta_5 \, P_{\Omega^\bot} \nabla_x\left(\dfrac{\rho^2}{2}\right) ,
$$
where
$$ \beta_5 =  \alpha \Phi_0 \, \Big(  \langle  \sin^2 \theta \, h \rangle_{M_\Om} +  \dfrac{1}{d(n-1)}  \langle  \sin^2 \theta \, \cos \theta \,  h \rangle_{M_\Om} \Big).$$
The terms $ {\mathcal T}_3 (\rho M_\Om)$ and $T_3$ have been computed in \cite{Degond_etal_MAA13}. In particular, it is easy to see that we get them from the formulae for  $ {\mathcal T}_2 (\rho M_\Om)$ and $T_2$ by changing  $- \alpha \Phi_0 \nabla_x (\rho^2/2)$ into $k_0 P_{\Omega^\bot} \Delta_x (\rho \Omega)$. Therefore, we get: 
$$
T_3 = \beta_6  \, P_{\Omega^\bot} \Delta_x(\rho \Om) ,
$$
where
 $$ \beta_6 = - k_0 \, \Big(  \langle  \sin^2 \theta \, h \rangle_{M_\Om} +  \dfrac{1}{d(n-1)}  \langle  \sin^2 \theta \, \cos \theta \,  h \rangle_{M_\Om} \Big).
 $$
 Inserting the expressions of $T_1, T_2$ and $T_3$ into (\ref{eq2}) we get (\ref{macro2}). \endproof

\section{Proof of Proposition \ref{prop1}  }
\label{numerical:convergence}

We follow the lines of the proof of Proposition 3.1 of \cite{Motsch_Navoret_MMS11}.
Assume that $\rho^\eta \rightarrow \rho^0$ and $\Om^\eta \rightarrow \Om^0 $ as $\eta$ tends to zero. Then, set
$$
R^\eta:=\rho^\eta (1 - |\Om^\eta|^2) \Om^\eta .
$$
Multiplying  equation (\ref{relax2}) by $\eta$ and then taking the limit $\eta \rightarrow 0$ yields $R^\eta \rightarrow 0.$ It follows that $|\Om^{0}|^{2} = 1.$ Since the vector $R^\eta$ is parallel to $\Om^\eta, $ we have $P_{(\Omega^\eta)^\bot}R^\eta = 0,$ which implies  that
$$
P_{(\Omega^\eta)^\bot}  \Big( \partial_t(\rho^\eta \Om^\eta)  +   \nabla_x \cdot  ( \rho^\eta V^\eta \otimes \Om^\eta)  +   \nabla_x  p(\rho^\eta)  -  \gamma \Delta_x(\rho^\eta \Om^\eta) \Big)= 0.
$$
Therefore, letting  $\eta \rightarrow 0,$ we obtain
\begin{eqnarray}
 & &  \partial_t(\rho^0 \Om^0) +     \nabla_x \cdot  ( \rho^0 V^0 \otimes \Om^0)  +     \nabla_x p( \rho^0)  -  \gamma \Delta_x(\rho^0 \Om^0) = \beta \Om^0,
 \label{relax33}
\end{eqnarray}
where $\beta$ is a real number, $p(\rho^\eta) \rightarrow  p( \rho^0) = d  \rho^0 + \alpha \Phi_0 (d + c_2) (\rho^0)^2/2 ,$ $V^0 =  c_2 \Om^0  -\Phi_0 \nabla_x \rho^0 $ and $U^0=c_1 \Om^0-\Phi_0 \nabla_x \rho^0$. By taking the scalar product of (\ref{relax33}) with $\Om^0$, we get
$$
\beta = \partial_t \rho^0 +  \nabla_x \cdot (\rho^0 V^0) +  \nabla_x p( \rho^0)\cdot\Om^0 -  \gamma  \Delta_x(\rho^0 \Om^0)\cdot \Om^0.
$$
Inserting this expression of $\beta$ into (\ref{relax33}) we find the equation for the evolution of the average direction (\ref{macro2}) and thus the SOHR model (\ref{macro1})-(\ref{macro22}). \endproof

\end{appendices}

\end{document}